\documentclass[a4paper,11pt]{article}
\pdfoutput=1 

\usepackage{jheppub} 

\usepackage[T1]{fontenc} 

\usepackage{placeins}
\usepackage{comment}
\usepackage{graphicx}
\usepackage{subfig}
\usepackage{cleveref}

\title{\boldmath Neutralino Dark Matter in Scenarios with Early Matter
Domination}

\author[a]{Manuel Drees,}
\author[a,b]{Fazlollah Hajkarim}

\affiliation[a]{Bethe Center for Theoretical Physics and Physikalisches
  Institut, Universit\"at Bonn,\\Nussallee~12, D-53115 Bonn,
  Germany}
  
\affiliation[b]{Institut f\"ur Theoretische Physik, Goethe Universit\"at 
Frankfurt, Max von Laue Strasse 1, D-60438 Frankfurt, Germany}
  
\emailAdd{drees@th.physik.uni-bonn.de}
\emailAdd{hajkarim@th.physik.uni-frankfurt.de}

\abstract{We investigate the production of neutralino dark matter in a
  cosmological scenario featuring an early matter dominated era ending
  at a relatively low reheating temperature. In such scenarios
  different production mechanisms of weakly interacting massive
  particles (WIMPs), besides the well--studied thermal production, can
  be important. This opens up new regions of parameter space where the
  lightest neutralino, as the best--known supersymmetric (SUSY) WIMP,
  obtains the required relic abundance. Many of these new sets of
  parameters are also compatible with current limits from colliders as
  well as direct and indirect WIMP searches. In particular, in
  standard cosmology bino--like neutralinos, which emerge naturally as
  lightest neutralino in many models, can have the desired relic
  density only in some finetuned regions of parameter space where the
  effective annihilation cross section is enhanced by co--annihilation
  or an $s-$channel pole. In contrast, if the energy density of the
  universe was dominated by long--lived PeV--scale particles
  (e.g. moduli or Polonyi fields), bino--like neutralinos can obtain
  the required relic density over wide regions of supersymmetric
  parameter space. We identify the interesting
    ranges of mass and decay properties of the heavy long--lived
    particles, carefully treating the evolution of the temperature of
    the thermal background.}

\begin{document} 
\def\gsim{\:\raisebox{-0.5ex}{$\stackrel{\textstyle>}{\sim}$}\:}
\def\lsim{\:\raisebox{-0.5ex}{$\stackrel{\textstyle<}{\sim}$}\:}

\maketitle
\flushbottom
\section{Introduction}
\label{intro}

The lightest neutralino as lightest supersymmetric particle (LSP) is
one of the oldest and most studied examples of a weakly interacting
massive particle (WIMP) candidate for the cosmological Dark Matter
(DM); see e.g.  \cite{Ellis:1983ew} for an early exploration of
parameter space, and \cite{Jungman:1995df,Roszkowski:2017nbc} for
reviews. The minimal supersymmetric extension of the Standard Model
(MSSM) contains four neutralino current eigenstates: a bino, a wino,
and two higgsinos.  Given current collider constraints on
superparticles, in particular on the masses of charginos and the
heavier neutralinos, we now know that over most of parameter space,
the mass eigenstates are relatively pure states, with little mixing.

Most analyses of WIMP DM worked in the framework of standard
cosmology, where the Universe was radiation--dominated starting at the
end of inflation and ending at a temperature around $1$ eV. Moreover,
it is usually assumed that the post--inflationary reheat temperature
was sufficiently high that WIMPs attained full thermal (chemical and
kinetic) equilibrium. The WIMP relic density is then basically
inversely proportional to its (effective) annihilation cross section
\cite{Kolb:1990vq, Gondolo:1990dk}. In that case higgsino--like WIMPs
typically need to have a mass near $1$ TeV to have the correct relic
density, and a wino--like WIMP should be at least two times
heavier. While it has recently been pointed out that these values
might be lowered by $30\%$ or so due to co--annihilation effects
\cite{Chakraborti:2017dpu}, the required values are still
uncomfortably high when compared to estimates of weak--scale
finetuning in the MSSM. In particular, while bounds on the masses of
scalar tops and gluinos based on simple loop calculations
\cite{Papucci:2011wy} are somewhat controversial
\cite{Feng:1999mn,Kitano:2006gv,Baer:2014ica}, it is generally agreed
that higgsino, and hence LSP, masses above several hundred GeV would
lead to percent level (or worse) finetuning; note that in the MSSM the
higgsino mass enters the relevant finetuning condition already at
tree--level.\footnote{This argument can be evaded \cite{Ross:2016pml}
  if there is a soft supersymmetry breaking contribution to the
  higgsino mass; this would not contribute to the Higgs boson masses
  which in turn determine finetuning. While this is technically
  possible, it would require a rather complicated supersymmetry
  breaking scenario.} In standard cosmology, higgsino-- or wino--like
WIMPs with masses in the few hundred GeV range would have too small a
relic density. In contrast, a bino--like WIMP has too large a relic
density in such a scenario, unless its effective annihilation cross
section is boosted by co--annihilation \cite{Griest:1990kh,
  Ellis:1998kh, Boehm:1999bj} or by an $s-$channel pole
\cite{Griest:1990kh, Drees:1992am}.

A predicted underdensity of WIMP DM can be cured by adding another DM
component, e.g. axions \cite{Tegmark:2005dy,Baer:2011hx,Baer:2014eja};
this can be done within the framework of minimal cosmology, and without
changing TeV--scale particle physics. On the other hand, a scenario
that predicts too large a relic density for a given DM candidate is
clearly excluded. This argument thus disfavors bino--like WIMPs, at
least within minimal cosmology.

At the same time bino--like WIMPs quite easily satisfy the
increasingly stringent constraints from direct WIMP searches
\cite{pdg,Aprile:2018dbl}; these searches exclude many scenarios where
the WIMP is higgsino--like, if the latter contributes most or all of
DM. Moreover, indirect searches \cite{pdg,Ahnen:2016qkx} now exclude
models where most or all of DM consists of wino--like (higgsino--like)
WIMPs with mass below $\sim 0.8$ $(\sim 0.4)$ TeV, but hardly
constrain the parameter space if the LSP is bino--like. These null
results therefore favor bino--like WIMPs. At the same time bino--like
neutralinos often emerge as LSP in simple models where the
superparticle spectrum can be described by a small number of free
parameters. In particular, if gaugino masses unify at or near the same
scale where the gauge couplings meet in the MSSM, the weak--scale bino
mass will be about half of the wino mass. Moreover, if stop squarks
and Higgs bosons have similar soft breaking masses at this very high
energy scale, the weak--scale higgsino mass parameter typically comes
out larger than the bino mass.

These arguments motivate us to investigate a non--minimal cosmological
scenario, in the hope of finding an extended region of parameter space
where a bino--like WIMP obtains the required relic density. In
particular, we analyse scenarios featuring an early matter--dominated
epoch sometime between the end of inflation and Big Bang
nucleosynthesis (BBN). This is quite well motivated, since
UV--complete theories like supergravity \cite{Polonyi:1977pj} and
superstring theory often contain heavy but long--lived scalar
particles, nowadays usually called moduli. They are long--lived since
their couplings to MSSM fields are suppressed by the inverse of the
Planck mass. Nevertheless they can attain large densities if their
mass is below the Hubble parameter during inflation
\cite{Vilenkin:1982wt,Linde:1982uu,Starobinsky:1982ee,
  Goncharov:1984qm,Dine:1995uk}. The success of standard BBN implies
that the moduli--dominated epoch should end at a final reheat
temperature of at least $4$ MeV \cite{Polnarev:1982,Coughlan:1983ci,
  Kawasaki:2000en,Hannestad:2004px, deSalas:2015glj}.

It has been pointed out more than ten years ago that the WIMP relic
density in this scenario can be either smaller or larger than in
standard cosmology \cite{Gelmini:2006pq,Gelmini:2006pw}. On the one
hand, since the moduli decay out of equilibrium, they increase the
entropy density of the universe, thereby diluting a pre--existing
WIMP density. At the same time new WIMP production mechanisms become
possible, including direct moduli to WIMP decays. In addition to the
effective WIMP annihilation cross section and mass, the final WIMP
relic density now also depends on the mass and lifetime of the modulus
as well as on the effective branching ratio for modulus to WIMP decays.

The impact of an early matter dominated epoch on WIMP DM has been
studied before \cite{Chung:1998rq, Moroi:1999zb, Giudice:2000ex,
  Pallis:2004yy, Gelmini:2006pq, Gelmini:2006pw, Acharya:2009zt,
  Arcadi:2011ev, Kane:2015qea,Hamdan:2017psw,Bernal:2018kcw}; specifically supersymmetric WIMPs were
considered in this context in \cite{Easther:2013nga,Allahverdi:2013noa,
  Roszkowski:2014lga, Aparicio:2015sda, Aparicio:2016qqb}. However, as
we pointed out recently \cite{Drees:2017iod}, an accurate treatment of
the radiation component during the decay of the moduli is very
important: even though the entropy is no longer conserved in this
scenario, one still has to 
normalize the WIMP number density to 
the known radiation density, which is related to the entropy density
; any inaccuracies in the calculation of the latter
therefore immediately affect the final prediction of the WIMP relic
density. Note also that the temperature of the thermal background
during modulus domination can vary over several orders of
magnitude. An accurate treatment of the temperature dependence of the
effective number of degrees of freedom therefore becomes mandatory if
one aims for precise predictions. Here we use the treatment of
ref.\cite{Drees:2015exa}, which in turn is based on lattice QCD
analyses of the equation of state, which determines the relation
between temperature and entropy and energy density in QCD. Moreover,
we use the latest results from direct and indirect WIMP searches to
further constrain the allowed parameter space. We find that large
regions of parameter space with good bino--like WIMP indeed survive,
if moduli are not too heavy and have a sufficiently small branching
ratio for decays into WIMPs.

The remainder of this article is organized as follows. In
Sec.~\ref{sec:friedmann-boltzmann} we discuss the set of equations we
need to solve in order to compute the WIMP relic density. Then
Sec.~\ref{sec:neutralino-thermal} briefly reviews neutralino DM in
standard cosmology, including a discussion of current experimental
constraints. In Sec.~\ref{sec:neutralino-nonthermal} we return to
moduli cosmology, delineating regions in parameter space where various
WIMP production mechanisms are important. We then present numerical
results from a random scan over the MSSM parameter space. Finally, we
summarize our results in Sec. \ref{sec:conclusion}.

\section{Basic Framework}
\label{sec:friedmann-boltzmann}

As stated in the Introduction, we use the formalism of
ref.\cite{Drees:2017iod}, which carefully treats the evolution of the
radiation component with temperature, using \cite{Drees:2015exa} for
the precise evolution of the effective number of degrees of freedom with
temperature.

In this article we ignore the details of thermalization process
\cite{Harigaya:2013vwa,Harigaya:2014waa,Mukaida:2015ria} by working in
the approximation that the decay products of the massive ``modulus''
particle $\phi$ (other than the WIMP $\chi$) thermalize
instantaneously. In particular, we do not consider the production of
WIMPs from reactions involving energetic $\phi$ decay products prior
to their thermalization, which can be important for WIMP masses not
too much smaller than the $\phi$ mass \cite{Allahverdi:2002pu}. In
this approximation we only need to consider the integrated Boltzmann
equations (for densities in normal space, rather than for phase space
densities) describing the evolution of the mass density $\rho_\phi$ of
the massive, very weakly interacting particle $\phi$, of the entropy
density $s_R$ of the radiation component, and of the WIMP number
density $n_\chi$:
\begin{eqnarray} \label{densities}
\frac {d\,\rho_{\phi}} {dt} + 3 H\rho_{\phi} &=& - \Gamma_{\phi}\rho_{\phi}\,, 
\nonumber \\
\frac {d\, s_R} {dt} + 3 Hs_R &=& \frac{1}{T} \left[ (1 - \bar{B}) \Gamma_{\phi }
 \rho_{\phi} + 2 \left<E\right>_{\textrm{eff}} \langle \sigma v \rangle_{\textrm{eff}} 
\left( n_{\chi}^2 - {n_{\chi,\rm EQ}}^2\right) \right]\,, \nonumber \\
\frac {d\, n_{\chi}} {dt} + 3 Hn_{\chi} &=& \frac{B_{\chi}}{M_{\phi}} \Gamma_{\phi} 
\rho_{\phi} - \langle \sigma v \rangle_{\textrm{eff}} \left( n_\chi^2 - 
n_{\chi,\rm EQ}^2\right)\,.
\end{eqnarray}
The first eq.(\ref{densities}) describes the reduction of the $\phi$
density through Hubble expansion and $\phi$ decay, $\Gamma_\phi$ being
the total $\phi$ decay width which is the inverse of the lifetime
$\tau_\phi$. The second equation describes the evolution of the
thermal background, whose entropy density is diluted by the Hubble
expansion but increased by $\phi$ decay into radiation as well as by
out--of--equilibrium annihilation of WIMPs; however, this last term is
always negligible in practice. The entropy production from $\phi$ decay
depends on the effective branching ratio
\begin{equation} \label{eq:Bbar}
\bar{B} = \frac {\langle E \rangle_{\textrm{eff}} B_\chi} {M_\phi}\,,
\end{equation}
with
$\langle E \rangle_{\textrm{eff}} \simeq \sqrt{M_\chi^2 + 3 T^2} \sim M_\chi$
for $T \ll M_\chi$, $B_\chi$ the average number of $\chi$
particles produced per $\phi$ decay and $M_\phi$ the mass of
$\phi$. Since we are interested in scenarios with $M_\chi \ll M_\phi$,
$\bar B \ll 1$ in all cases of interest.\footnote{Frequently an
  equation for the evolution of the energy density of radiation is
  used instead of the one for $s_R$. However, as emphasized in
  \cite{Kolb:1990vq} this is only correct if the number of degrees of
  freedom is constant during $\phi$ decay, which is usually not the
  case in the scenarios we consider.} Finally, the last
eq.(\ref{densities}) describes the evolution of the WIMP number
density, which is again diluted by Hubble expansion and enhanced by
$\phi \rightarrow \chi$ decays; the last term of this equation
describes annihilation of WIMPs into SM particles as well as inverse
annihilation of SM particles, assumed to be in thermal equilibrium,
into WIMPs.

The Hubble parameter appearing in eqs.(\ref{densities}) is as usual
determined by the Friedmann equation:
\begin{equation} \label{friedmann}
H^2 = \frac {8 \pi \rho_{\rm tot}} {3 M_{\rm Pl}^2} = 
\frac { 8 \pi  \left( \rho_\phi + \rho_R+ \rho_\chi \right)} {3 M_{\rm Pl}^2}\,,
\end{equation}
where $M_{\rm Pl} \simeq 1.22 \cdot 10^{19}$ GeV is the Planck mass.
Since the $\chi$ number density becomes comoving constant well within
the radiation--dominated epoch after $\phi$ decay, the last term in
eq.(\ref{friedmann}) is very small for the range of temperatures over
which we need to solve the Boltzmann equations numerically, but the
contribution from radiation is important; it is given by
\begin{equation}\label{rho_R}
\rho_{R}(T) = \frac{\pi^2}{30} g_*(T) T^4\,,
\end{equation} 
$g_{*}$ being the number of relativistic degrees of freedom defined
via $\rho_R$.  The temperature $T$ is computed from the entropy
density, which is given by
\begin{equation} \label{eq:s_R}
s_R (T) = \frac{\rho_R(T)+p_R(T)}{T} = \frac{2\pi^2}{45} h_*(T) T^3\,,
\end{equation}
where $h_{*}$ is another measure for the number of relativistic degrees of
freedom. A massless, non--interacting particle contributes equally to $g_*$
and $h_*$, but particle masses as well as the strong interactions around
the QCD deconfinement transition affect $g_*$ and $h_*$ differently 
\cite{Drees:2015exa}.

As mentioned in the Introduction, the relic density of the lightest
neutralino is often affected by co--annihilation effects. In
particular, if the LSP is wino--like, its mass is very similar to that
of the lightest chargino; a higgsino--like LSP is close in mass to
both the lightest chargino and the second--lightest neutralino. We
include co--annihilation using the framework developed in
ref.\cite{Griest:1990kh}. Note that reactions turning the LSP into one
of its co--annihilation partners or vice versa remain in equilibrium
well after the system of superparticles decouples from the SM
particles. Moreover, after this decoupling eventually each
superparticle will decay into an LSP [plus some irrelevant SM
particle(s)]. These observations allow to interpret $n_\chi$ appearing
in eqs.(\ref{densities}) as the sum over the number densities of {\em
  all} (relevant) superparticles, $n_\chi = \sum_{i}n_i$, and
$n_{\textrm{eq}} = \sum_i {n_{\textrm{eq},i}}$ for the equilibrium
densities. The effective thermally averaged cross section for
superparticle annihilation can then be written as
\cite{Edsjo:1997bg,Roszkowski:2014lga}
\begin{equation} \label{eq:sigeff}
\langle\sigma v\rangle_{\textrm{eff}} = \sum_{i=1} \sum_{j=1} 
{\langle\sigma_{ij}v_{ij}\rangle\, \frac{n_{\textrm{eq},i}} {n_{\textrm{eq}}} 
\frac{n_{\textrm{eq},j}}{n_{\textrm{eq}}}}\,.
\end{equation}
Here the double sum runs over all superparticles that can
co--annihilate with the LSP. In practice it is sufficient to include
weakly interacting particles whose masses are within about $20\%$ of
the LSP mass, whereas strongly interacting superparticles within $30$
or $35\%$ of the LSP mass should be considered.

The modulus dominated epoch ends when most $\phi$ particles have
decayed. Here we are interested in scenarios with a fairly long early
matter dominated era. Today's radiation then almost all comes from
$\phi$ decays. In the instantaneous decay approximation, the
``reheat'' temperature after $\phi$ decay is given by
\begin{equation} \label{eq:T_RH}
T_{\rm RH} = \sqrt{\Gamma_\phi M_{\rm Pl}} \left( \frac {45} {4 \pi^3 g_*(T_{\rm RH})}
 \right)^{1/4}\,.
\end{equation}
In practice this equation has to be solved iteratively, due to the
temperature dependence of $g_*$ appearing on the right--hand side (rhs).
Since we treat $\phi$ decays exactly, rather than in the instantaneous
decay approximation, $T_{\rm RH}$ strictly speaking has no physical meaning;
it nevertheless remains a good measure characterizing the end of the
early matter dominated epoch.

The only free parameter appearing in eq.(\ref{eq:T_RH}) is the $\phi$
decay width. We make the usual assumption that its coupling to (some)
(MS)SM particles are suppressed by a single power of the Planck mass,
in which case
\begin{equation} \label{decaywidth}
\Gamma_\phi = \alpha \frac{M_\phi^3} {M_{\rm Pl}^2}\,, \ \ 
\alpha=\frac{C}{8 \pi}=~{\rm constant}\,.
\end{equation}
The value of the constant $C$ is model dependent. In the rest of this
paper we assume $\alpha=1$. In scenarios with sufficiently long early
matter domination, the final result depends essentially only on
$\Gamma_\phi$ and $B_\chi/M_\phi$ once all densities have been
normalized to today's radiation density. Results for $\alpha \neq 1$
can thus be read off from our numerical results by choosing a
different value of $M_\phi$ such that $\Gamma_\phi$ is the same as in
our ansatz, and then rescaling $B_\chi$ so that the original value of
$B_\chi/M_\phi$ is restored; this will work as long as the resulting
$M_\phi \gg M_\chi$ and the resulting $B_\chi < 1$.

For our numerical work we follow ref.\cite{Arcadi:2011ev} and introduce
dimensionless quantities:
\begin{equation} \label{eq:densities}
\Phi \equiv  \frac{\rho_{\phi} A^3}{T_{\rm RH}^4}\,, \,\, 
R \equiv \rho_R \frac{A^4} {T_{\rm RH}^4}\,, \,\, 
X \equiv n_{\chi} \frac{A^3}{T_{\rm RH}^3}\,, \,\,
A \equiv a T_{\rm RH}\,,
\end{equation}
where $a$ is the scale factor in the Friedmann--Robertson--Walker metric.
The Hubble parameter can then be written as
\begin{equation} \label{eq:hubble_2}
H= \widetilde{H} T_{\rm RH}^2 A^{-3/2} c_1^{-1/2} M_{\rm Pl}^{-1} \,,
\end{equation}
where $c_1=\frac{3}{8 \pi}$ and the dimensionless Hubble parameter
$\tilde{H}$ is given by
\begin{equation} \label{eq:hubble_1}
\widetilde{H} \equiv \left(\Phi + \frac{R}{A} + 
\frac{\langle E\rangle_{\textrm{eff} }  X }{T_{\rm RH}} \right)^{1/2}\,.
\end{equation}

We also use the dimensionless quantity $A$ to parameterize the
evolution of the Universe, rather than the (cosmological) time $t$
appearing in the original Boltzmann equations (\ref{densities}). The
new, dimensionless evolution equations are thus:\footnote{It turns out
  to be numerically more convenient to use $\ln(A)$ as evolution
  parameter, rather than $A$ itself, with $A \ d/dA = d/d\, \ln(A)$.}
\begin{eqnarray} \label{eq:boltzmann}
\widetilde{H} \frac{d\Phi}{dA} &=& -\,c_{\rho}^{1/2}\,A^{1/2} \Phi \, ; 
\nonumber \\ 
\widetilde{H} \frac{d X}{d A } &=& \frac{c_\rho^{1/2} T_{\rm RH} B_{\chi}}
{M_{\phi}} A^{1/2} \Phi + c_1^{1/2}\, M_{\rm pl} T_{\rm RH} A^{-5/2} \, 
\langle \sigma v \rangle _{\textrm{eff}}\left( {X_{\rm EQ}}^2 - {X}^2 \right)\, ;
\nonumber \\
\frac{dT}{dA} &=& \left( 1 + \frac{T}{3 h_*} \frac{dh_*}{dT} 
\right)^{-1}  \bigg[ -\frac{T}{A} + \frac {15 T_{\rm RH}^6} 
{2 \pi^2 c_1^{1/2} M_{\rm Pl} H T^3 h_* A^{\frac{11}{2}} } 
\big( c_{\rho}^{1/2}\,A^{3/2} (1 - \bar{B})  \Phi 
\nonumber \\  && \hspace*{30mm} 
+ c_1^{1/2}\, M_{\rm pl}\, \frac{2 \left<E\right>_{\textrm{eff}} 
\langle \sigma v \rangle _{\textrm{eff}}} {A^{3/2}} \left({X}^2 - 
{X_{\rm EQ}}^2 \right) \bigg]\,.
\end{eqnarray}
Here we have introduced $c_{\rho}=\frac{\pi^2 g_{*}(T_{RH})}{30}$. 

We solve these differential equations with the initial conditions
\cite{Giudice:2000ex,Kane:2015qea,Drees:2017iod}
\begin{equation} \label{init}
A_I = 1, \,\,\,
X_I = 0, \,\,\, 
R_I =0, \,\,\, 
\Phi_I = \frac{3H_I^2M_{Pl}^2}{8\pi T_{RH}^4},\,\,\, 
H_I=\gamma \Gamma_{\phi}\,.
\end{equation}
The case with non--vanishing initial WIMP and radiation densities has
been studied in \cite{Drees:2017iod}. There we found that an initial
$R_I \leq \Phi_I$ does not change the result if
$\gamma \gsim 10^{20}$. The following argument shows that this is not
difficult to achieve. The primordial modulus field, created during
inflation, begins to behave like an ensemble of free particles at
$H = M_\phi$. The earliest plausible $H_I$ is thus given by $M_\phi$.
$H_I \gsim 10^{20} \Gamma_\phi$ therefore requires
$M_\phi \gsim 10^{20} \Gamma_\phi$, which via eq.(\ref{decaywidth})
implies $M_\phi \lsim 10^{-10} M_{\rm Pl}$. On the other hand, the
requirement that WIMPs do not attain thermal equilibrium after $\phi$
decay requires $T_{\rm RH} \lsim M_\chi / 20$, and hence via
eq.(\ref{eq:T_RH}),
$M_\phi \lsim 8 \cdot 10^7 \ {\rm GeV} \left[ M_\chi / (1 \ {\rm TeV})
\right]^{2/3}$,
which is a significantly stronger bound for $M_\chi \leq 1$
TeV. Moreover, at $H = M_\phi$ the universe will only have been
radiation--dominated if at that time the temperature exceeded
$2 \cdot 10^{12} \ {\rm GeV} \left[ M_\phi / (10^7 \ {\rm GeV})
\right]^{1/2}$.
If the temperature at $H = M_\phi$ was below this value, the universe
was $\phi$ matter dominated for all $H$ between $M_\phi$ and
$\Gamma_\phi$, and the initial radiation content can be safely
ignored; and even if the temperature at $H = M_\phi$ was somewhat
above this (already quite high) value, the modulus dominated epoch
might still have been long enough to make the initial radiation, and
$\chi$, content irrelevant.

As mentioned above, the final DM relic density is obtained by normalizing
to the known radiation density \cite{Drees:2017iod}:
\begin{eqnarray} \label{relicmod}
\Omega_{\chi} h^2 &=& \frac{ \rho_{\chi}(T_{\rm now})} {\rho_\gamma(T_{\rm now})} 
\Omega_\gamma h^2
= \frac{\rho_{\chi}(T_E)} {2 \rho_R(T_E)} \frac {g_*(T_E)
h_*(T_{\rm now})} {h_*(T_E)} \frac{T_E} {T_{\rm now}} \Omega_{\gamma} h^2 
\nonumber \\ 
&=& M_{\chi} \frac {X(T_E)} {R(T_E)} \frac {A_E T_E g_*(T_E) 
h_*(T_{\rm now})} {2 T_{\rm now} T_{\rm RH} h_*(T_E)} \Omega_\gamma h^2 \, .
\end{eqnarray}
Here $\Omega_i$ denotes today's mass or energy density of component
$i$ in units of the critical density (i.e., $\Omega_{\rm tot} = 1$
yields a flat universe), $h$ is the Hubble constant in units of
$100 \ {\rm km} / ({\rm Mpc} \cdot {\rm s})$, and $T_{\rm now}$ is the
current temperature of the cosmic microwave background (CMB)
radiation. $A_E$ is the value of $A$ where the numerical solution of
the Boltzmann equations is terminated. It should be well within the
radiation--dominated epoch after all $\phi$ particles have decayed
and $\chi$ has completely decoupled, so that $R$ and $X$ become
constant. Note that the entropy density is 
comoving-constant for $A > A_E$.
Finally, we use the PDG values for today's quantities
\cite{pdg}:
\begin{equation} \label{CMB}
\Omega_\gamma h^2 = 2.473\times10^{-5}\,, 
~T_{\rm now} = 2.7255~{\rm K} = 2.35\times10^{-13}\ {\rm GeV}\,.
\end{equation}
The present DM relic density is also quite well known
\cite{pdg}\footnote{The value given in eq.(\ref{DMden}) has been
  obtained within the minimal cosmological model. However, it should
  be valid in our framework as well, since an early matter dominated
  epoch does not affect the evolution of density perturbations that
  are probed by current cosmological observations, in particular by
  the CMB anisotropies. Moreover, the precise value of the DM relic
  density is immaterial for the arguments of this paper.}
\begin{equation} \label{DMden}
\Omega_{DM}h^2= 0.1186 \pm 0.002 \, .
\end{equation}
Since we normalize to today's radiation density, the final result is
essentially independent of the initial $\phi$ density $\Phi_I$, as
long as the period of early matter domination lasted sufficiently long
($\gamma \gsim 10^{10}$) and $R_I, X_I \ll 1$. The reason is that in this
limit {\em all} densities are proportional to $\Phi_I$, which therefore
drops out in the ratio used in eq.(\ref{relicmod}).

\section{Thermal Neutralino Dark Matter}

\label{sec:neutralino-thermal}

In this Section we briefly review neutralino DM in standard cosmology, and
describe our numerical procedures.

In the MSSM with $R-$parity conservation \cite{Martin:1997ns,book} the
lightest supersymmetric particle is stable; it can
be a neutralino which is therefore a potential DM candidate
\cite{Ellis:1983ew}. The MSSM contains four neutralino interaction (or
current) eigenstates: a bino, a wino, and two higgsinos, where the
former are superpartners of the neutral $U(1)_Y$ and $SU(2)$ gauge
bosons and the latter are superpartners of the two MSSM Higgs
doublets. Upon electroweak symmetry breaking these states mix via a
$4 \times 4$ mass matrix. However, over most of parameter space the
mixing is rather small, i.e. usually the four mass eigenstates are
dominated by a single current eigenstate, with small admixtures of the
three other states. This is true because searches at LEP and the LHC
require \cite{pdg} the wino and higgsino mass parameters to be
significantly above the mass of the $W$ bosons, which sets the scale
for the off--diagonal entries in the neutralino mass matrix. The
neutralino masses are therefore essentially set by the soft breaking
bino mass $M_1$, the soft breaking wino mass $M_2$ and the
supersymmetric higgsino mass $\mu$. Note that the bino is a gauge
singlet. $M_1$ is therefore still essentially unconstrained
experimentally, if one does not assume it to be related to $M_2$
\cite{Dreiner:2009ic}.

If $|M_1| < |M_2|\,, |\mu|$ the LSP will be bino--like. In most of parameter
space its annihilation cross section is dominated by the exchange of
sfermions in the $t-$ and $u-$channel \cite{Drees:1992am}, so that
\begin{equation} \label{binosig}
\langle \sigma_{\tilde{B}} v\rangle \propto \frac {\alpha_{\rm em}^2}
{\cos^4 \theta_W} \sum_{\tilde f} \frac {Y_{\tilde f}^4 M_1^2} {m_{\tilde f}^4}
\frac {M_1}{T}\,
\end{equation}
where the last factor is due to the $P-$wave suppression of the
annihilation of a Majorana fermion into massless SM fermions; here
$\alpha_{\rm em}$ is the fine structure constant and $\theta_W$ is the
weak mixing angle. Recall that in standard cosmology, the WIMP relic
density is inversely proportional to its (effective) annihilation
cross section \cite{Kolb:1990vq}. The increasing lower bounds on the
sfermion masses \cite{pdg} imply that the cross section
(\ref{binosig}) is too small, leading to bino overdensity in standard
cosmology. This can only be evaded if the bino co--annihilates,
e.g. with a $\tilde \tau$ slepton \cite{Ellis:1998kh} or a $\tilde t$
squark \cite{Boehm:1999bj}, or if the annihilation cross section is
enhanced by a nearby resonance, in particular for $M_1 \simeq m_A/2$
where $m_A$ is the mass of the neutral CP--odd Higgs boson of the MSSM
\cite{Drees:1992am}

For $|\mu| < |M_1|\,, |M_2|$ the LSP is higgsino--like and can annihilate
efficiently into $W^+ W^-$ and $Z^0Z^0$ pairs, via the exchange of other
higgsino--like chargino and neutralino states, respectively. In this case
annihilation from an $S-$wave initial state is allowed, leading to a
thermally averaged cross section \cite{ArkaniHamed:2006mb}
\begin{equation} \label{higgsinosig}
\langle \sigma_{\tilde{H}} v\rangle
=\frac{g^4}{512 \pi \mu^2} \left(21+3 \tan^2 \theta_W+11\tan^4\theta_W\right)\,.
\end{equation}
The presence of these other higgsino--like states, whose masses are
within a few GeV of that of the LSP, implies that
co--annihilation effects are always important for higgsino--like LSP
\cite{Mizuta:1992qp,Edsjo:1997bg}. Here also other final states
contribute (e.g. $W^\pm Z^0$ for $\tilde \chi_1^\pm \tilde \chi_1^0$
co--annihilation, or $f \bar f$ pairs via $Z^0$ exchange for
$\tilde \chi_1^0 \tilde \chi_2^0$ co--annihilation, where $f$ is an SM
fermion). All these cross sections also scale like $1/\mu^2$, so that
in standard cosmology
\begin{equation} \label{omegahiggsino}
\Omega_{\tilde{H}} h^2  \simeq 0.10 \left( \frac{\mu}{1 ~{\rm TeV}} \right)^2\,.
\end{equation}
Note that Sommerfeld enhancement is quite small for higgsino--like LSP
\cite{Drees:2013er}.

Finally, for  $|M_2| < |M_1|\,, |\mu|$ the LSP is wino--like, with thermally
averaged cross section \cite{ArkaniHamed:2006mb}
\begin{equation} \label{winosig}
\langle \sigma_{\tilde{W}}v\rangle= \frac{3g^4}{16\pi M_2^2},
\end{equation}
As in the case of higgsino--like LSP, co--annihilation is always important
for wino--like LSP, since the charged wino is very close in mass to the
neutral one. The effective cross section determining the wino relic density
in standard cosmology therefore differs somewhat from eq.(\ref{winosig}),
but it still satisfies
\begin{equation} \label{omegawino}
\Omega_{\tilde{W}} h^2 \propto M_2^2 \,.
\end{equation}
Even in the absence of Sommerfeld enhancement \cite{Hisano:2003ec,
  Hisano:2006nn} a wino--mass near $2.5$ TeV is required in order to
obtain the desired relic density (\ref{DMden}) in standard
cosmology. Note also that a wino--like LSP is not compatible with the
unification of gaugino masses near the scale where the MSSM gauge
couplings unify; in such models $M_2 \simeq 2 M_1$ at the weak scale.

In standard cosmology there are no moduli fields, and the comoving
entropy density is conserved during WIMP decoupling. We therefore only
have to solve the second eq.(\ref{eq:boltzmann}). We used {\tt
  MicrOMEGAs} \cite{Belanger:2001fz} for this purpose;\footnote{In
  {\tt MicrOMEGAs} the thermally averaged cross section is a function
  of temperature $T$. Even in nonstandard cosmology WIMPs will be in
  thermal equilibrium until the temperature falls well below the WIMP
  mass. This function is thus only needed for $T \ll M_\chi$. In our
  numerical calculation we therefore set
  $\langle \sigma v \rangle (T\geq M_\chi/13) =\langle \sigma v
  \rangle (T=M_\chi/13)$,
  keeping the full temperature dependence only for $T < M_\chi/13$.
  This helps to speed up the calculation without
  significant loss of accuracy in the final result.}
this program also allows to compute the LSP annihilation cross section
in today's universe, needed to check indirect detection constraints,
and the LSP--nucleon scattering cross sections,
which are constrained by direct detection experiments. Moreover, we
used {\tt SuSpect } \cite{Djouadi:2002ze} for the computation of the
spectrum of superparticles and Higgs bosons in the MSSM. The scans
over parameter space where performed with the help of the {\tt T3PS}
\cite{Maurer:2015gva} which parallelizes the computation, leading to
great gains in speed. This is not crucial for standard cosmology, but
becomes important in the non--standard scenario, where the solution of
the Boltzmann equations becomes considerably more computationally
intensive.

We performed two different scans. The first is in the framework of the
phenomenological MSSM (pMSSM), where the soft breaking parameters are
defined directly at the TeV scale. Here we follow
ref. \cite{Roszkowski:2014lga} and parameterize the spectrum with 10
independent free parameters (``p10MSSM''): the masses of the three
MSSM gauginos $M_1, \, M_2, \, M_3$; the trilinear scalar soft
breaking parameters for $\tilde t$ squarks and $\tilde \tau$ sleptons,
$A_t, A_\tau$; the higgsino mass $\mu$; the mass $m_A$ of the CP--odd
neutral Higgs boson; a common soft breaking mass $m_{\widetilde Q_3}$
for third generation squarks; a common soft breaking mass
$m_{\widetilde L_3}$ for third generation sleptons; and the ratio of
vacuum expectation values $\tan\beta$. Since the $A$ parameters are
multiplied with the corresponding Yukawa couplings, they are
irrelevant for the first and second generation, so we set them to
zero. The $A$ parameter for ${\tilde b}$ squarks has
also been fixed; note that $L-R$ mixing in the
${\tilde b}$ sector is dominated by $\mu$, not by
$A_b$, since $\mu$ comes with a factor $\tan\beta$ here. The sfermion
masses of the first and second generation are chosen to lie
sufficiently far above $M_1$ that these sfermions do not contribute
significantly to co--annihilation; however, co--annihilation with
third generation sfermions is possible.

We vary all these parameters simultaneously in a random scan, using
flat distributions for the values between the limits shown in
table~\ref{pmssm}. We only consider points where the predicted mass
of the lighter CP--even Higgs boson lies within $3$ GeV of the measured
value of $125$ GeV \cite{Aad:2015zhl,Aaboud:2018wps}. Our scan probably
includes some points with first and second generation squark or gluino
masses below current bounds \cite{pdg}; however, the precise values of
these parameters basically do not affect the physics of neutralino DM,
apart from possible co--annihilation (which our lower limit of the squark
mass range excludes, as mentioned above). The lower limit of the scan
range of third generation squark masses is also quite low; however, the
requirement of a sufficiently heavy Higgs boson requires TeV--scale
stop masses, which satisfy current LHC search limits. Since in the
pMSSM scan $M_1, \, M_2$ and $\mu$ are varied independently, all three
types of LSP can occur.

\begin{table}[h]
\centering
\begin{tabular}{|c|c|}
\hline 
Parameter & Range \\ 
\hline
\hline 
bino mass & $0.1 < M_1 < 5$ \\ 
wino mass & $0.1 < M_2 < 6$ \\ 
gluino mass & $0.7 < M_3 < 10$ \\ 
stop trilinear coupling & $-12 < A_t < 12$ \\ 
stau trilinear coupling & $-12 < A_{\tau} < 12$ \\ 
sbottom trilinear coupling & $A_b = -0.5$ \\ 
CP--odd Higgs mass & $0.2 < m_A < 10$ \\ 
higgsino mass & $0.1 < \mu < 6$ \\ 
3rd gen. soft squark mass & $0.1 < m_{\widetilde{Q}_3} < 15$ \\ 
3rd gen. soft slepton mass & $0.1 < m_{\widetilde{L}_3} < 15$ \\ 
1st/2nd gen. soft squark mass  & $m_{\widetilde{Q}_{1,2}} = M_1 + 100$ GeV \\ 
1st/2nd gen. soft slepton mass  & $m_{\widetilde{L}_{1,2}} = m_{\widetilde{Q}_3} + 1$ TeV \\ 
ratio of Higgs doublet VEVs & $2 < \tan\beta < 62$ \\ 
\hline 
\end{tabular}
\caption{\small The parameters of the p10MSSM and their ranges used in our
  scan. All masses and trilinear couplings are given in TeV, unless
  indicated otherwise. All the parameters of the model are given at the
  superparticle mass scale. The range of parameters is similar to that chosen in ref.~\cite{Roszkowski:2014lga}.}
\label{pmssm} 
\end{table}

We also have performed another scan over the parameter space of the
constrained MSSM (cMSSM), where the soft breaking parameters are
assumed to unify near the scale of Grand Unification,
$M_X \simeq 2 \cdot 10^{16}$ GeV.  Specifically, at this very high
scale all gauginos are assumed to have the same mass $m_{1/2}$, and
all scalars (sfermions as well as Higgs bosons) get a common soft
breaking mass $m_0$. The trilinear soft breaking terms also unify to
$A_0$. $\tan\beta$ is again a free parameter. We chose the sign of
$\mu$ to be positive, as in the pMSSM scan; this has little effect on
the relic density, but slightly increases the spin--independent
scattering cross section on nucleons for bino--like LSP
\cite{Drees:1993bu}. The physical masses of superparticles and Higgs
bosons are again computed with the help of {\tt SuSpect}, which also
solves the relevant renormalization group equations. The range
of parameters we used is shown in table~\ref{cmssm}.

\begin{table}[t]
\centering
\begin{tabular}{|c|c|}
\hline 
Parameter & Range \\ 
\hline
\hline 
scalar mass & $0.1 < m_0 < 6$ \\ 
gaugino mass & $0.1 < m_{1/2} < 6$ \\ 
trilinear coupling & $-12 < A_0 < 12$ \\ 
ratio of Higgs doublet VEVs & $1 < \tan\beta < 60$ \\ 
sign of $\mu$ parameter & $\mu > 0$\\
\hline 
\end{tabular}
\caption{\small The range of cMSSM parameters we used in our scan are given at 
the GUT scale where the MSSM gauge couplings meet. All dimensionful parameters
 are given in TeV.} 
\label{cmssm} 
\end{table}

In the cMSSM the LSP is usually bino--like. Since gaugino masses are
assumed to unify, a wino--like LSP is not possible in this scenario.
Moreover, the large top Yukawa coupling in most cases drives the
squared soft breaking mass of one of the Higgs bosons to large
negative values, requiring a large value of $\mu$ in order to obtain the
correct mass of the $Z$ boson; in this model a higgsino--like LSP is possible 
only if $m_0 \gg m_{1/2}$. Moreover, co--annihilation with $\tilde t_1$
is also impossible in this framework, and an enhancement of the
bino annihilation cross section through resonant exchange of the heavy
Higgs bosons in the $s-$channel is possible only for very large values of
$\tan\beta$. 

\begin{figure}[h]
\centering
\includegraphics[width=75mm]{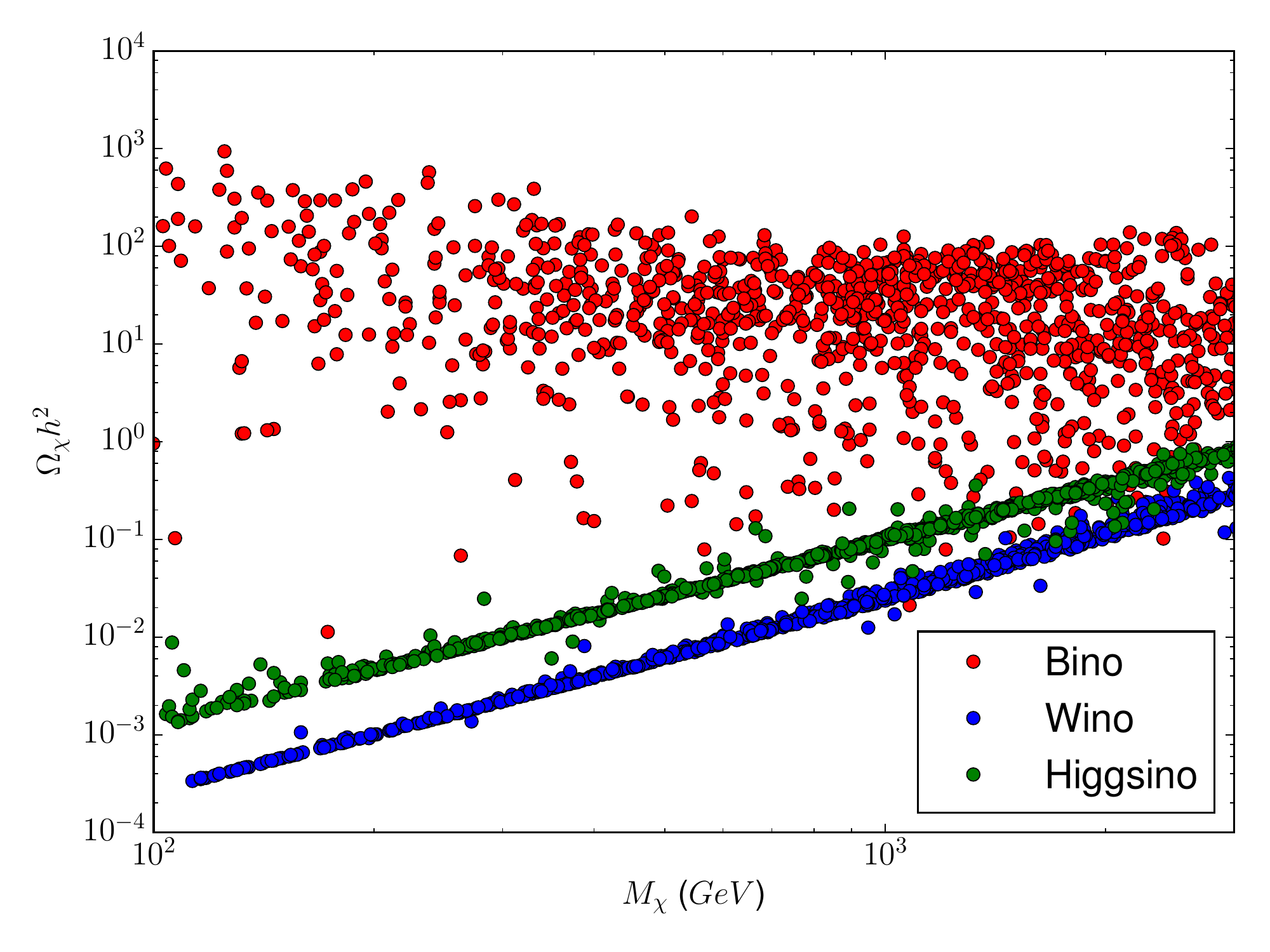}
\includegraphics[width=75mm]{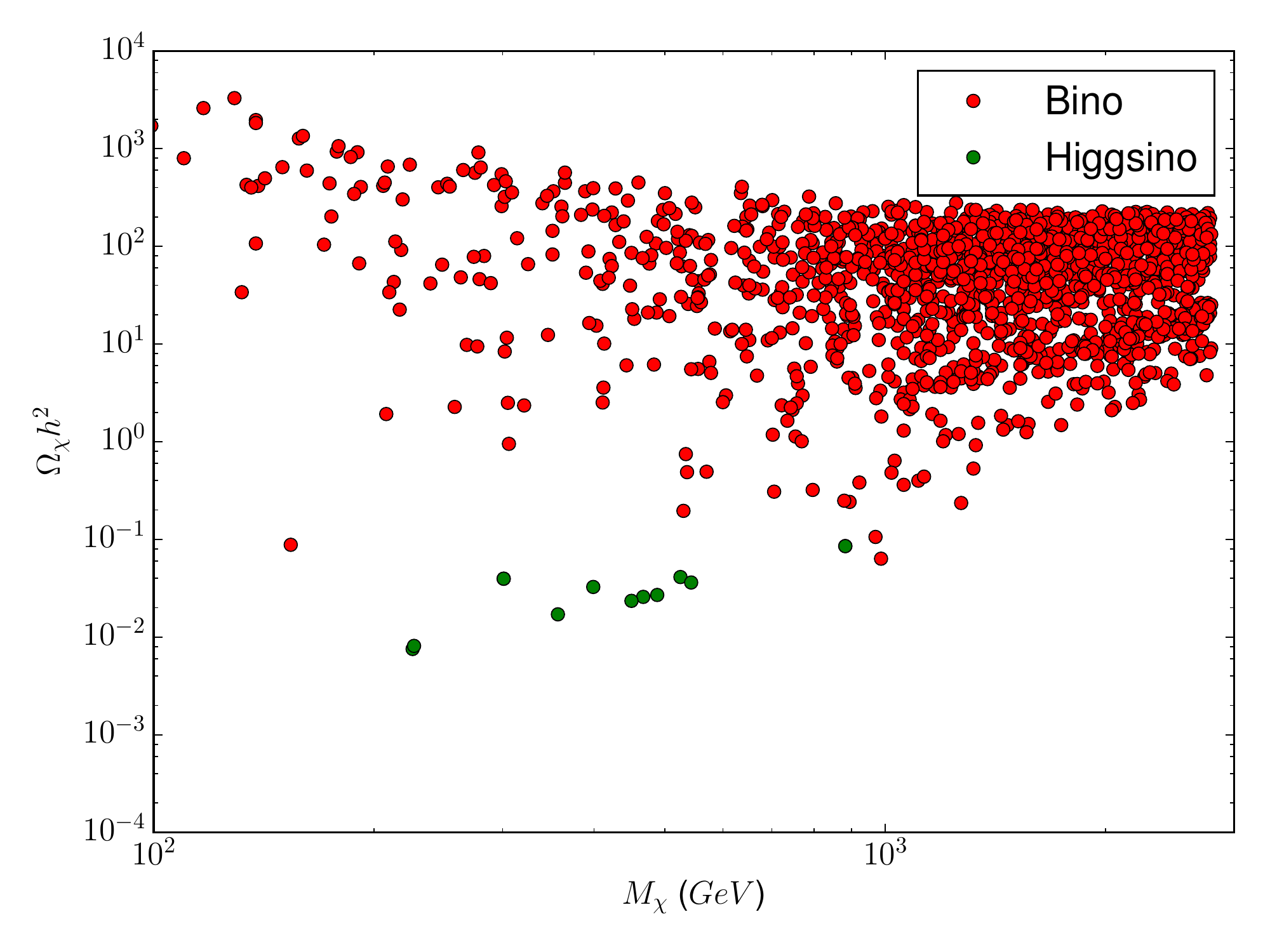}
\caption{Scatter plot of the predicted neutralino LSP relic density
in the p10MSSM (left) and the cMSSM (right) vs the mass of the neutralino. 
Red, green and blue points are for bino--, higgsino-- and wino--like LSP, 
respectively. Here we have assumed standard cosmology.
\label{thermal-relic-dm}}
\end{figure}

Examples of the predicted thermal relic density are shown in
Fig.~\ref{thermal-relic-dm}, for the p10MSSM (left) and cMSSM (right).
In the left frame we observe two bands of points with relic density
$\propto M_\chi^2$. Here the LSP is higgsino--like (green points) or
wino--like (blue points); the observed behavior conforms with the
expectations of eqs.(\ref{omegahiggsino}) and (\ref{omegawino}). In
contrast, the red points, where the LSP is bino--like, are widely
scattered, since the effective bino annihilation cross section depends
not only on the LSP mass, but also on sfermion masses; in addition,
$s-$channel Higgs exchange resonances can play a role. As already
noted in the Introduction, scenarios with bino--like LSP typically
lead to (much) too large a relic density, whereas higgsino-- or
wino--like LSPs attain the correct relic density only for masses that
require quite severe electroweak finetuning.

As expected, there are no points with wino--like LSP in the cMSSM scan
(right frame of fig.~\ref{thermal-relic-dm}), and very few points
where the LSP is higgsino--like. Moreover, the number of points with
rather light bino--like LSP is also relatively small. This is because
large gaugino, and hence bino, masses help to increase the weak--scale
stop masses; as noted above, quite large stop masses are needed in
order to reproduce the Higgs mass of about 125 GeV. Furthermore the
relic density of bino--like LSPs tends to be even higher than in the
pMSSM scan, due to the reduced opportunities for co--annihilation and
resonant $s-$channel enhancement, as explained above.

\begin{figure}[h]
\includegraphics[width=75mm]{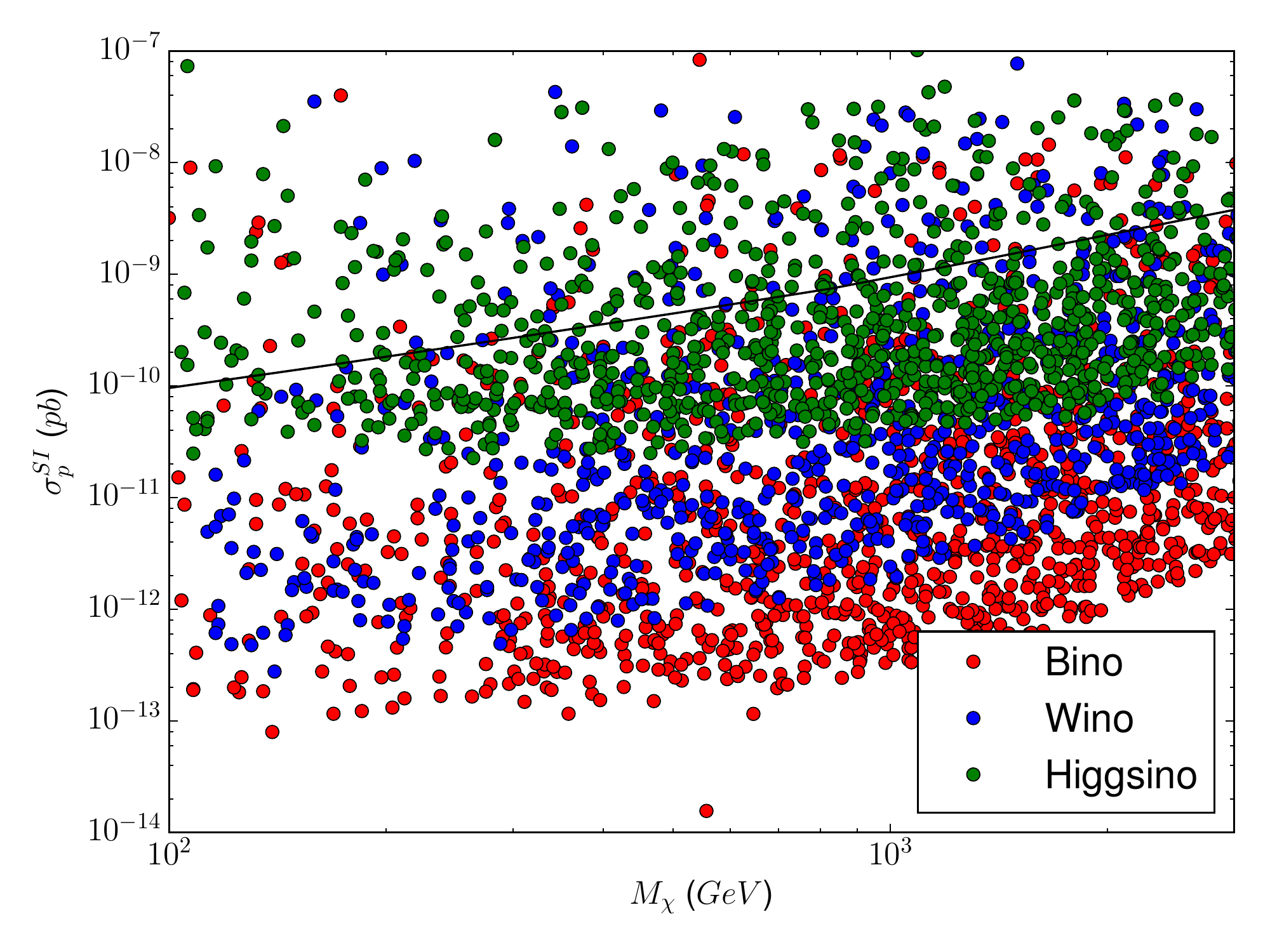}
\includegraphics[width=75mm]{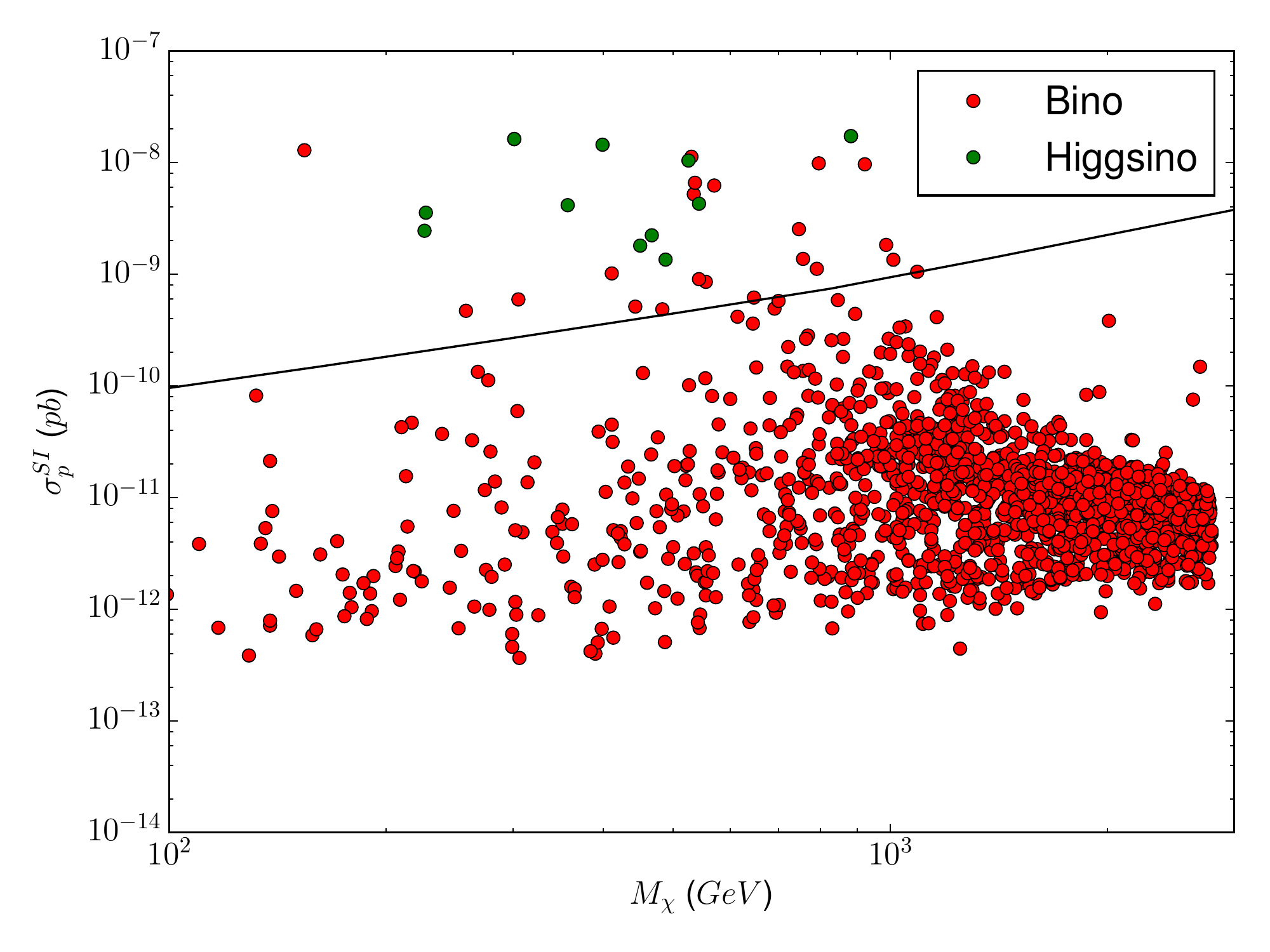}
\caption{The spin--independent LSP--proton scattering cross section
in the pMSSM (left) and in the cMSSM (right) vs the mass of the neutralino. 
The meaning of the colors is as in fig.~\ref{thermal-relic-dm}. Points 
above the diagonal line are excluded by the most recent Xenon1T bound 
\cite{Aprile:2018dbl}, if the neutralino LSP forms all of DM.
\label{direct-dm}}
\end{figure}

We now turn to the direct detection constraints depicted in
fig.~\ref{direct-dm}. Note that this constraint, as well as the
constraint from indirect detection discussed below, assume that the
given LSP forms all of dark matter (in our galaxy). We saw that this
is frequently not the case in standard cosmology. Since these
constraints do not directly depend on the cosmological scenario, we
nevertheless already discuss them here.

We see that most of the (red) points with bino--like LSP are safely below
the bound. In the left, pMSSM, frame the few red points above the bound
have quite small masses for first generation squark and/or the heavy neutral
Higgs boson; such scenarios are very difficult to realize in the cMSSM, so
in that scenario nearly all points with bino--like LSP survive this constraint.
In contrast, many points with higgsino--like LSP are excluded even in the
pMSSM case. In the cMSSM scan {\em all} points with higgsino--like
LSP are excluded. The reason is that in the cMSSM even for the points with
higgsino--like LSP the higgsino mass is not much smaller than that of
the bino. This leads to relatively large higgsino--bino mixing, which
yields relatively large couplings of the LSP to neutral Higgs bosons, and
hence rather large scattering cross sections. We finally note that most
points with wino--like LSP in the pMSSM scan survive this constraint even
if winos form all of DM.

\begin{figure}[h]
\includegraphics[width=75mm]{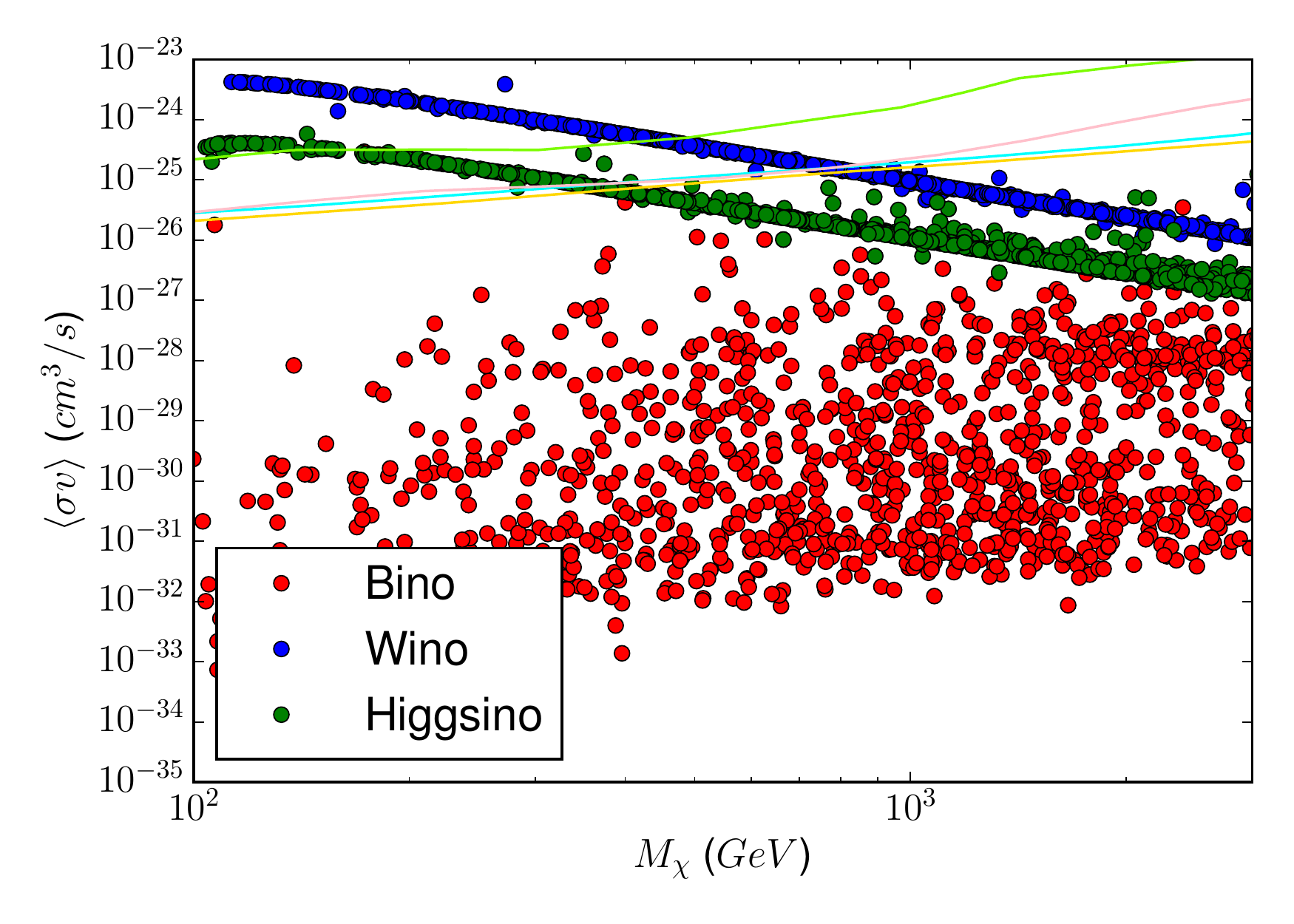}
\includegraphics[width=75mm]{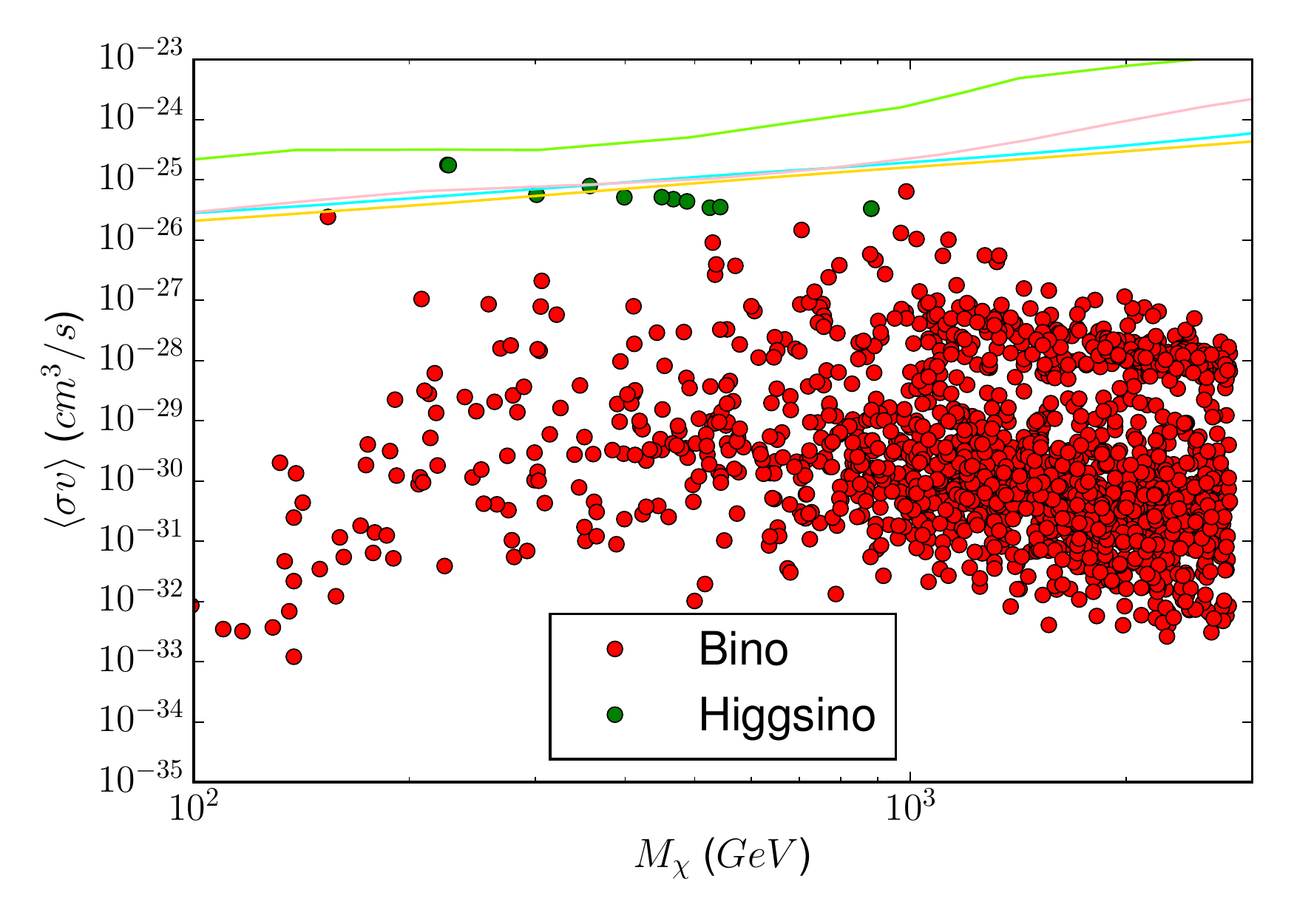}
\caption{The total $S-$wave neutralino annihilation cross section
in the pMSSM (left) and in the cMSSM (right) vs the mass of the neutralino. 
The meaning of the colors is as in fig.~\ref{thermal-relic-dm}. 
The cyan, yellow, pink, and light green lines are the constraints from 
the combination of MAGIC and Fermi--LAT results for the annihilation of 
Majorana DM to $W^+W^-$, $b^+b^-$, $\tau^+\tau^-$, and $\mu^+\mu^-$ final 
states, respectively \cite{Ahnen:2016qkx}. Points above these lines are
excluded if neutralinos form all of DM.
\label{indirect-dm}}
\end{figure}

Finally, the thermally averaged neutralino annihilation cross section
in today's universe is depicted in fig.~\ref{indirect-dm}. Since WIMPs
are now very non--relativistic, this is essentially the $S-$wave
contribution in the limit $v \rightarrow 0$. The solid lines are
exclusion contours from a combination of recent $\gamma$ telescope
data, for different assumptions of the dominant final state. These
bounds have again been obtained under the assumption
that a given WIMP forms all of DM.

We see that these constraints exclude wino--like neutralinos with
$M_\chi \leq 0.8$ TeV, and higgsino--like LSP with $M_\chi \leq 0.4$ TeV.
Note that co--annihilation does not play any role in today's universe.
Wino-- and higgsino--like LSPs both annihilate predominantly into heavy
gauge bosons, i.e. the relevant exclusion limit is the one shown by the cyan
curves. In contrast, all scenarios with bino--like LSP are allowed. The
safety margin is even higher in the cMSSM scan, since a large $s-$channel
resonance enhancement of the annihilation cross section is even less likely
there, as noted above.

Figures \ref{direct-dm} and \ref{indirect-dm} clearly favor bino--like
neutralinos, especially for the ``natural'' region of parameter space
where $M_\chi \lsim 0.5$ TeV; higgsino--like states near that mass are
also acceptable, if somewhat marginal in view of finetuning and
approaching constraints from both direct and indirect searches. On the
other hand, fig.~\ref{thermal-relic-dm} shows that in standard
cosmology, such binos usually have much too high a relic density, whereas
the relic density of $0.5$ TeV higgsinos is too low. This motivates the
investigation of our non--standard cosmological scenario, to which we
now return.

\section{Neutralino  Production in Non--Standard Cosmology}
\label{sec:neutralino-nonthermal}

As noted above, an early matter dominated epoch can
either increase or reduce the
predicted WIMP relic density, depending on the values of the free
parameters. The most relevant ones are the ratio of the ``reheating''
temperature $T_{\rm RH}$ at the end of that epoch, given by
eq.(\ref{eq:T_RH}); the mass $M_\chi$ of the WIMP; the WIMP
annihilation cross section, which (by crossing) also determines the
rate of WIMP production (or ``inverse annihilation''); and the
combination $M_\chi B_\chi / M_\phi$, which determines the importance
of direct $\phi \rightarrow \chi$ decays. In this section we first
make some general remarks on the parameter space, and then present
some examples of scenarios leading to the approximately correct relic
density for bino-- or higgsino--like neutralinos with ``natural''
masses.

\subsection{Discussion of the Parameter Space}

Here we analyse the dependence of various LSP production mechanisms on the
mass $M_\phi$, which determines the reheat temperature via eq.(\ref{eq:T_RH}),
and on the branching ratio for $\phi \rightarrow \chi$ decays.

\subsubsection{Modulus Mass}

Let us first discuss the relevant range of $M_\phi$. Its lower limit
is set by the requirement that modulus decay should not interfere with
big bang nucleosynthesis. This requires \cite{deSalas:2015glj} $T_{\rm
RH} \geq 4$ MeV, and hence via eqs.(\ref{eq:T_RH}) and (\ref{decaywidth}),
\begin{equation} \label{mphi_min}
M_\phi \geq 10^5 \ {\rm GeV}\,.
\end{equation}

On the other hand, if $\phi$ is very heavy, $T_{\rm RH}$ becomes so
large that WIMPs will maintain (or reach) full thermal equilibrium
even after all $\phi$ particles have decayed. This would bring us back
to standard cosmology as far as WIMP physics is concerned. Recalling that
in standard cosmology, WIMPs decouple at $T = T_F \simeq M_\chi/20$, and
taking into account that in reality the end of $\phi$ domination is not
sharp, we define the upper end of the interesting range of $M_\phi$ through
the condition $T_{\rm RH} \leq M_\chi /10$, which implies
\begin{equation} \label{mphi_max}
M_\phi \leq 2 \cdot 10^8 {\rm GeV} \cdot \left( \frac {M_\chi}
{1 \ {\rm TeV}} \right)^{2/3}\,.
\end{equation}
In both eqs.(\ref{mphi_min}) and (\ref{mphi_max}) we have again
assumed that the effective $\phi$ coupling parameter $\alpha =
1$. Eq.(\ref{decaywidth}) implies that both bounds on $M_\phi$ should be
multiplied by $\alpha^{-1/3}$ if $\alpha \neq 1$.

It is important to note that even for $\phi$ masses well below the
upper bound (\ref{mphi_max}) thermal contributions to the final LSP
relic density can be important. This is because $\phi$ decays quickly
create a thermal background \cite{Giudice:2000ex}, whose maximal
temperature exceeds $T_{\rm RH}$ by a factor $\gamma^{1/4}$
\cite{Drees:2017iod}. If the effective LSP annihilation cross section
is not too small, LSPs will therefore typically attain full thermal
equilibrium during the $\phi$ matter dominated epoch. The resulting
contribution to the final LSP relic density scales like
\cite{Giudice:2000ex}
\begin{equation} \label{mod_fo}
\Omega _{\chi,\, \rm ann} h^2[{\rm FO^{mod}}] \propto \frac { T_{\rm RH}^3 
{x_{\rm FO}}^4 }  {M_\chi^3 \langle \sigma v\rangle}\,,
\end{equation}
where $x_{\rm FO} = \frac{M_{\chi}}{T_{\rm FO}}$, $T_{\rm FO}$ being
the LSP decoupling temperature in the $\phi$ dominated epoch. Note
that this contribution to the relic density again scales like the
inverse of the effective LSP annihilation cross section, but the
coefficient is smaller than in standard cosmology by a factor
$\propto (T_{\rm RH}/T_{\rm FO})^3$. For a given LSP mass and
annihilation cross section, $T_{\rm FO}$ is somewhat larger than the
freeze--out temperature in standard cosmology, since for a given
temperature the Hubble parameter is larger by a factor
$(T/T_{\rm RH})^2$; this translates in a stronger $M_\chi$ dependence
of $x_{\rm FO} \equiv M_\chi / T_{\rm FO}$. Moreover, a decrease of
$T_{\rm RH}$ implies an increase of $T_{\rm FO}$, reducing the
suppression factor $(T_{\rm RH}/T_{\rm FO})^3$ even further.

For small $M_\phi$, and hence small $T_{\rm RH}$, and very small
annihilation cross section, LSPs may not attain thermal equilibrium at
all during the $\phi$ matter dominated epoch. In case of neutralino
LSPs, this may happen for $M_\phi \lsim 10^6$ GeV and $M_\chi \gsim 1$
TeV. Even in this case there will in general be a thermal contribution
to the final LSP relic density, due to the production of
superparticles from the thermal plasma.  As long as the LSP
annihilation cross section is below that required to obtain full
thermal equilibrium, this contribution will be proportional to this
cross section \cite{Giudice:2000ex}. Finally, independent of whether
LSPs attain thermal equilibrium during $\phi$ domination or not, the
thermal contribution to the LSP relic density is bounded from above
\cite{Giudice:2000ex}:
\begin{equation} \label{max_th}
\Omega_{\chi, \, {\rm thermal}} h^2\lsim 0.1 \left( \frac{T_{\rm RH}} {1 \ {\rm GeV}}
\right)^5 \left( \frac {100 \ {\rm GeV}} {M_\chi} \right)^4 \,.
\end{equation}
For our choice $\alpha = 1$ this means that thermal LSP production can
become relevant only for
$M_\phi \gsim 3 \cdot 10^6 \ {\rm GeV} \left( 100 \ {\rm GeV} / M_\chi
\right)^{8/15}$.
Note that the upper bound (\ref{max_th}) can get saturated only for
rather small annihilation cross section; for higgsino-- or wino--like
neutralinos, which attained full thermal equilibrium during $\phi$
domination for all cases of interest, the maximal thermal contribution
is significantly smaller.

\subsubsection{Branching Fraction}

In addition to thermal production, neutralinos can also be produced 
from $\phi$ decays. Note that here not only direct decays into the LSP are
relevant, but decays into {\em all} superparticles, since the heavier
superparticles will quickly decay into the lightest neutralino. 

Moreover, most $\phi$ particles decay at a temperature near
$T_{\rm RH}$. The upper bound (\ref{mphi_max}) ensures that $\chi$ particles
are not in thermal equilibrium at that temperature.\footnote{Of course,
for $M_\phi$ just below this upper bound, thermal effects at $T \simeq
T_{\rm RH}$ will not be completely negligible.} The final LSP relic density
can then be written as \cite{Kane:2015qea}
\begin{equation} \label{sum}
\Omega_\chi h^2 = \Omega_{\chi, \, {\rm thermal}} h^2+ \Omega_{\chi, \, {\rm decay}} h^2\,,
\end{equation}
where $\Omega_{\chi, \, {\rm thermal}}h^2$ originates from $\chi$ interactions with
the thermal plasma and has been discussed above. Here we analyze the second
contribution, which is due to $\phi \rightarrow \chi$ decays.

Although by assumption $\chi$ particles were not in thermal
equilibrium at $T \simeq T_{\rm RH}$, the $\chi$ density may attain
``quasi--static'' equilibrium, where the rhs of the second
eq.(\ref{eq:boltzmann}) vanishes due to a cancellation between the
term $\propto X^2$ and the term $\propto B_\chi$. This will happen if
the LSP annihilation cross section is above a critical value
\cite{Kane:2015qea}, which scales like
$M_\phi / (B_\chi M_{\rm Pl} T_{\rm RH}^2)$.  Using
eqs.(\ref{eq:T_RH}) and (\ref{decaywidth}) and writing
$\langle \sigma v \rangle = \kappa /M_\chi^2$ where $\kappa$ is
dimensionless, this can be translated into a critical value of
$B_\chi$:
\begin{equation} \label{B_crit}
B_{\chi, \, {\rm crit}} = \frac {8\pi} {3\alpha} \frac{1}{\kappa}
\left( \frac {M_\chi} {M_\phi} \right)^2
\simeq \left\{ \begin{array}{l}
4 \cdot 10^3 \left( M_\chi / M_\phi \right)^2\,, \ \ \chi \simeq \tilde H\,; \\
7 \cdot 10^2 \left( M_\chi / M_\phi \right)^2\,, \ \ \chi \simeq \tilde W\,.
\end{array} \right.
\end{equation}
In the second step we have given approximate numerical values for higgsino--
and wino--like LSP. The annihilation cross section of bino--like LSPs does
not simply scale like $M_\chi^{-2}$ times a constant; in fact, in most cases
of interest the bino annihilation cross section is below the critical value.

If the annihilation cross section is well above its critical value or,
equivalently, if $B_\chi > B_{\chi, \, {\rm crit}}$,
$\Omega_{\chi}h^2$ is independent of $B_\chi$ and scales inversely
with the effective annihilation cross section \cite{Kane:2015qea}. The
dependence on $\langle \sigma v \rangle$ is thus the same as in
standard cosmology, but the relic density is parametrically enhanced
by a factor $T_{\rm F} / T_{\rm RH}$, where $T_F \simeq M_\chi/20$ is
the LSP decoupling temperature in standard cosmology. Schematically,
\begin{equation} \label{decay1}
\Omega_{\chi, \, {\rm decay}} h^2 \propto \frac {M_\chi} {\langle \sigma v \rangle
T_{\rm RH}}\,.
\end{equation}

In the opposite limit, where the annihilation cross section is well below
its critical value, $\Omega_{\chi, \, {\rm decay}}h^2$ is simply proportional to
$B_\chi$:
\begin{equation} \label{decay2}
\Omega_{\chi, \, {\rm decay}} h^2 \propto \frac {B_\chi T_{\rm RH} M_{\chi}}
{M_\phi} \propto B_\chi M_\chi M_\phi^{1/2}\,.
\end{equation}
The factor $T_{\rm RH}/M_\phi$ results because the $\phi$ number density
$n_\phi = \rho_\phi / M_\phi$, and $\rho_\phi$ is related to the entropy
density $s$ after $\phi$ decay via $\rho_\phi \propto T_{\rm RH} s$, with
$s$ being removed by the final normalization to today's photon density.
In the second step we have used $T_{\rm RH} \propto M_\phi^{3/2}$, see
eqs.(\ref{eq:T_RH}) and (\ref{decaywidth}).

\subsection{Numerical Results}

We are now ready to present some numerical results yielding
cosmologically interesting relic densities for neutralinos, in
particular for relatively light bino-- or higgsino--like neutralinos;
we saw in Sec.~3 that indirect searches already exclude wino--like
neutralinos as main contributor to the total DM density for
``natural'' masses, which we interpret as $M_\chi \lsim 500$ GeV. We use
the same MSSM parameters as in Sec.~3. The resulting relic density can
therefore be directly compared to Fig.~\ref{thermal-relic-dm}. We use
the same computer codes as for the case of standard cosmology, including
some custom--written new routines for {\tt MicrOMEGAs} that solve the
coupled system of Boltzmann equations (\ref{eq:boltzmann}).

We structure this discussion by analyzing different values for
$M_\phi$ within the range defined by eqs.(\ref{mphi_min}) and
(\ref{mphi_max}).

\subsubsection{Light Moduli}

We first discuss the case $M_\phi = 5 \cdot 10^5$ GeV, corresponding
to $T_{\rm RH} = 40.68$ MeV; recall that we assume $\alpha = 1$ in
eq.(\ref{decaywidth}). This scenario is rather simple to analyse. On
the one hand, the bound (\ref{max_th}) implies that neutralino
production from the thermal bath is negligible in this
case.\footnote{For $M_\phi < 10^6$ GeV and very small $B_\chi$,
  thermal production -- in particular, the ``inverse annihilation'' of
  SM particles into pairs of LSPs -- can give the dominant
  contribution to the LSP relic density. However, this is still
  ``negligible'' in the sense that the resulting DM density is several
  orders of magnitude smaller than the desired value (\ref{DMden}).}
Moreover, eq.(\ref{B_crit}) shows that the annihilation of neutralinos
produced in $\phi$ decay become significant only for
$B_\chi \gsim 10^{-3}$ even for wino--like LSP.

\begin{figure}[h]
\includegraphics[width=75mm]{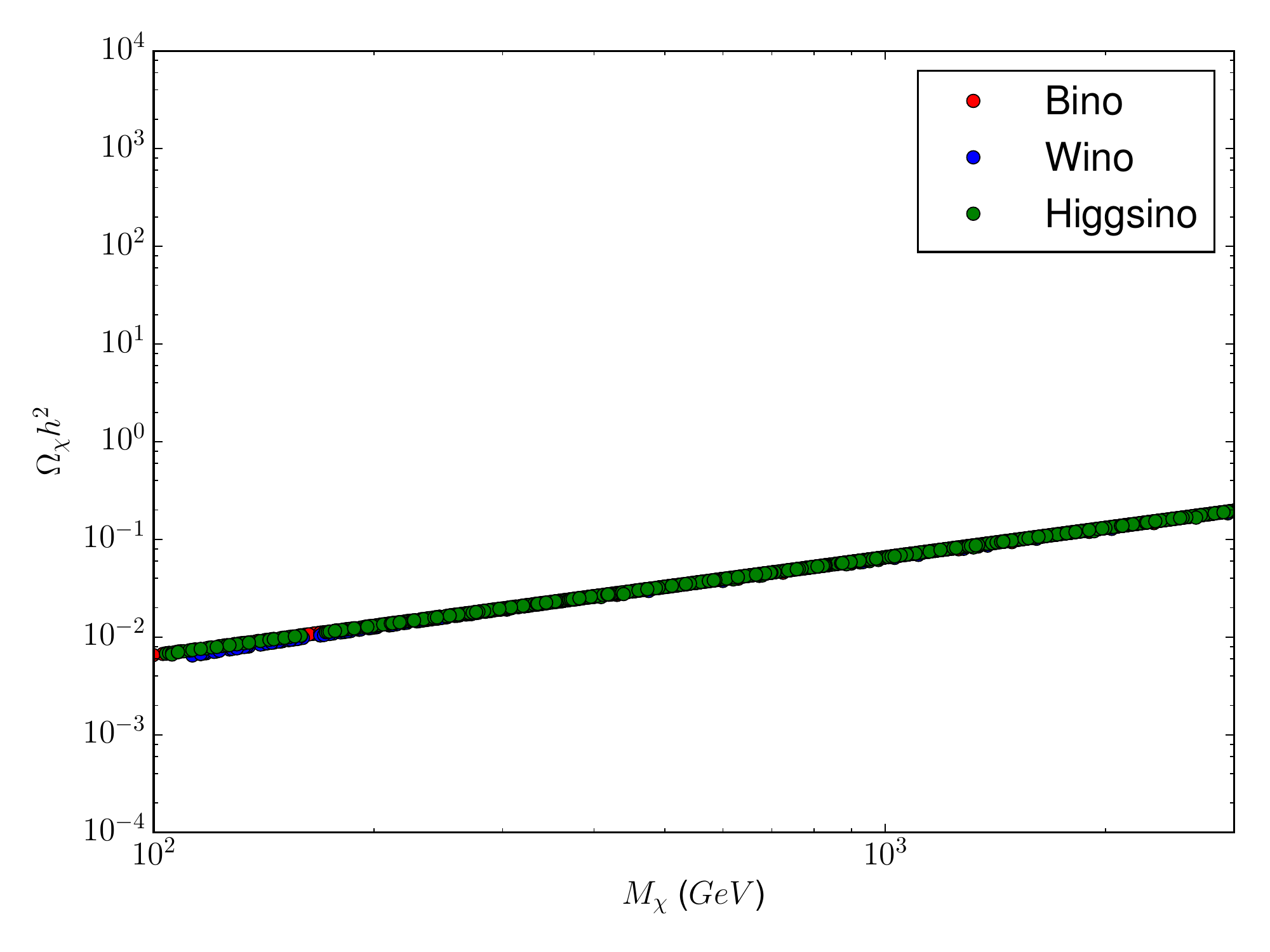}
\includegraphics[width=75mm]{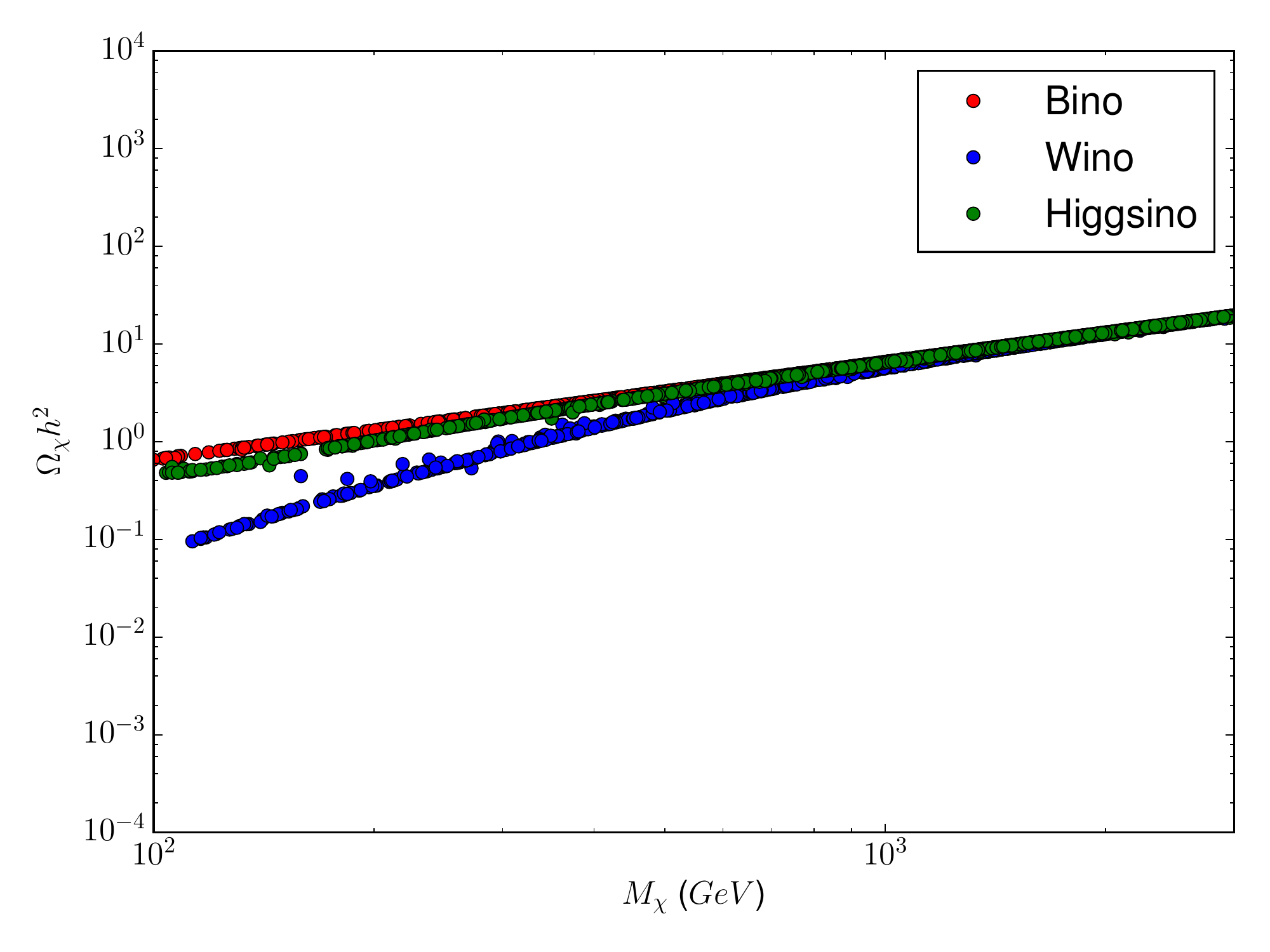}
\caption{The LSP relic density vs LSP mass for $M_\phi = 5 \cdot 10^5$
  GeV in the pMSSM, for $B_\chi = 10^{-5}$ (left) and $10^{-3}$
  (right). Red, green and blue points again stand for bino--,
  higgsino-- and wino--like LSP, respectively. These results can be
compared with the left frame of Fig.~\ref{thermal-relic-dm}.
\label{5e5-pmssm}}
\end{figure}

These considerations are confirmed by Fig.~\ref{5e5-pmssm}. In the
left frame $B_{\chi} = 10^{-5}$. In that case neutralino annihilation is
always negligible, hence the result is independent of
$\langle \sigma v \rangle$. On the other hand, for $B_\chi = 10^{-3}$
(right frame) the annihilation of wino--like neutralinos becomes
relevant for $M_\chi < 1$ TeV, but it is still basically irrelevant for
bino-- and higgsino-- like neutralinos. 

This allows to determine the first choice of parameters giving the
desired relic density for ``natural'' bino-- or higgsino--like LSP:
\begin{equation} \label{range_1}
B_\chi \simeq 1.5 \cdot 10^{-4} \cdot \frac {100 \ \rm GeV} {M_\chi} \cdot
\left( \frac {5 \cdot 10^5 \ {\rm GeV}} {M_\phi} \right)^{1/2}\,.
\end{equation}
This equation is quite accurate for $M_\phi = 5 \cdot 10^{5}$ GeV for
which the results of Fig.~\ref{5e5-pmssm} have been obtained. It also
works fairly well for other values in the range
$10^5
\leq M_\phi \leq 10^6$ GeV, where the lower bound had been given in
eq.(\ref{mphi_min}). For $M_\phi \gsim 10^6$ GeV, $T_{\rm  RH}$ comes 
close to or exceeds the QCD transition temperature, so that the effective 
numbers of degree of freedom $g_*$ and $h_*$ show a sizable dependence 
on $T_{\rm  RH}$ and hence on $M_\phi$, which is not captured by the simple 
expression (\ref{range_1}). For slightly larger $M_\phi$
contributions to the final LSP relic density from interactions with
the thermal plasma can also become significant.

Nevertheless eq.(\ref{range_1}) remains a reasonable approximation
even for $M_\phi \lsim 10^7$ GeV as long as the annihilation cross
section is sufficiently small, e.g. for bino-- or higgsino--like
neutralinos with $M_\chi \gsim 300$ GeV. Moreover, it works for all
versions of the MSSM, and in fact for all WIMPs whose annihilation
cross section does not substantially exceed that of wino--like
neutralinos, simply because WIMP annihilation is negligible in these
cases. The only difference in the cMSSM is that a wino--like LSP
cannot be realized, and a higgsino--like LSP is rare, as already
discussed in Sec.~3. We therefore do not show numerical results for
the cMSSM in this light moduli scenario.

\subsubsection{Intermediate--Mass Moduli}

We now turn to the case $M_\phi = 5 \cdot 10^6$ GeV, corresponding to
$T_{\rm RH} = 848.60$ MeV. This case is more complicated than the
previous scenario, since now thermal and non--thermal LSP production
can both be important; annihilation of non--thermally produced
neutralinos also plays a prominent role for some range of parameters
leading to a cosmologically interesting DM relic density.

\begin{figure}[h]
\includegraphics[width=75mm]{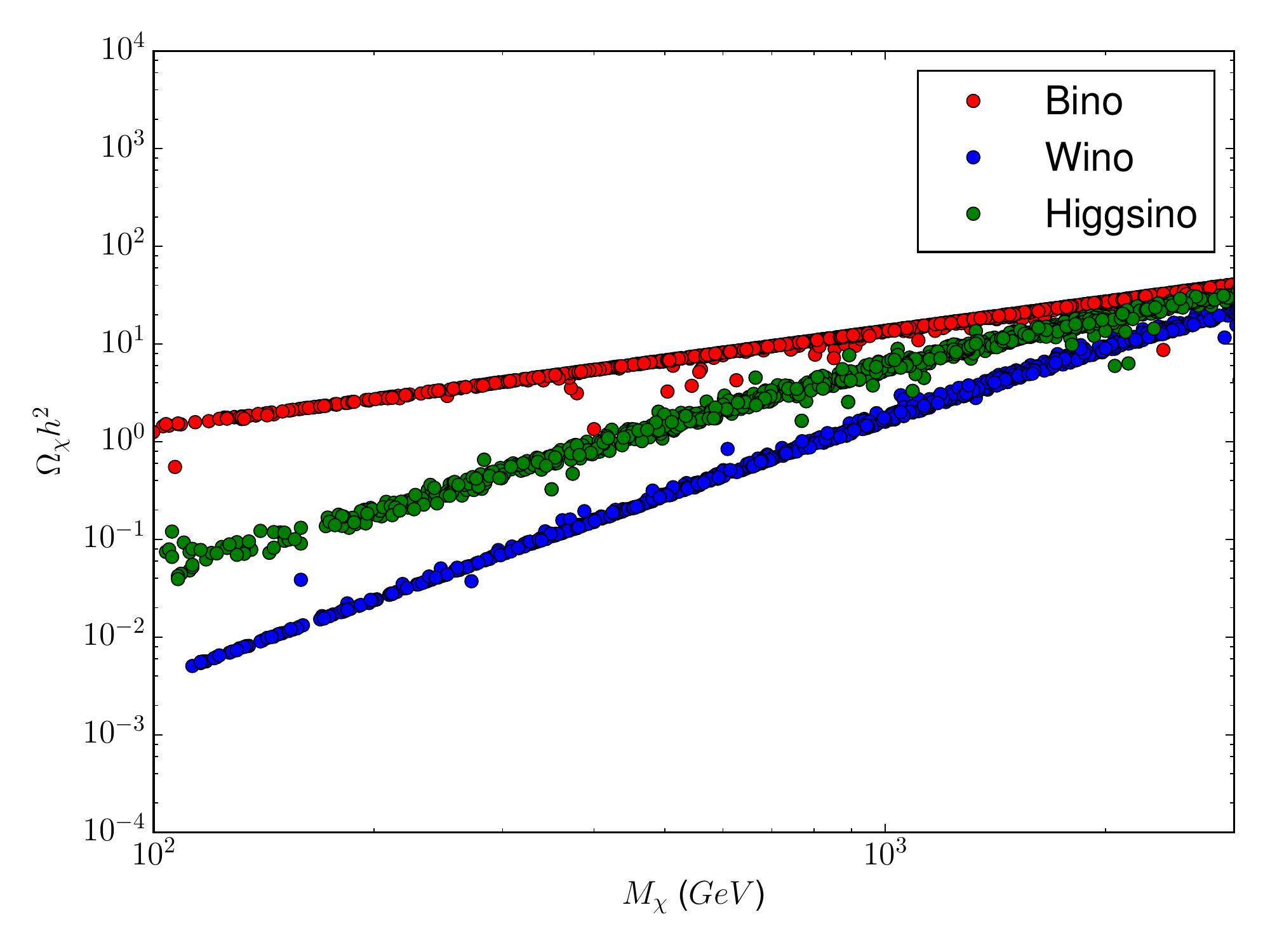}
\includegraphics[width=75mm]{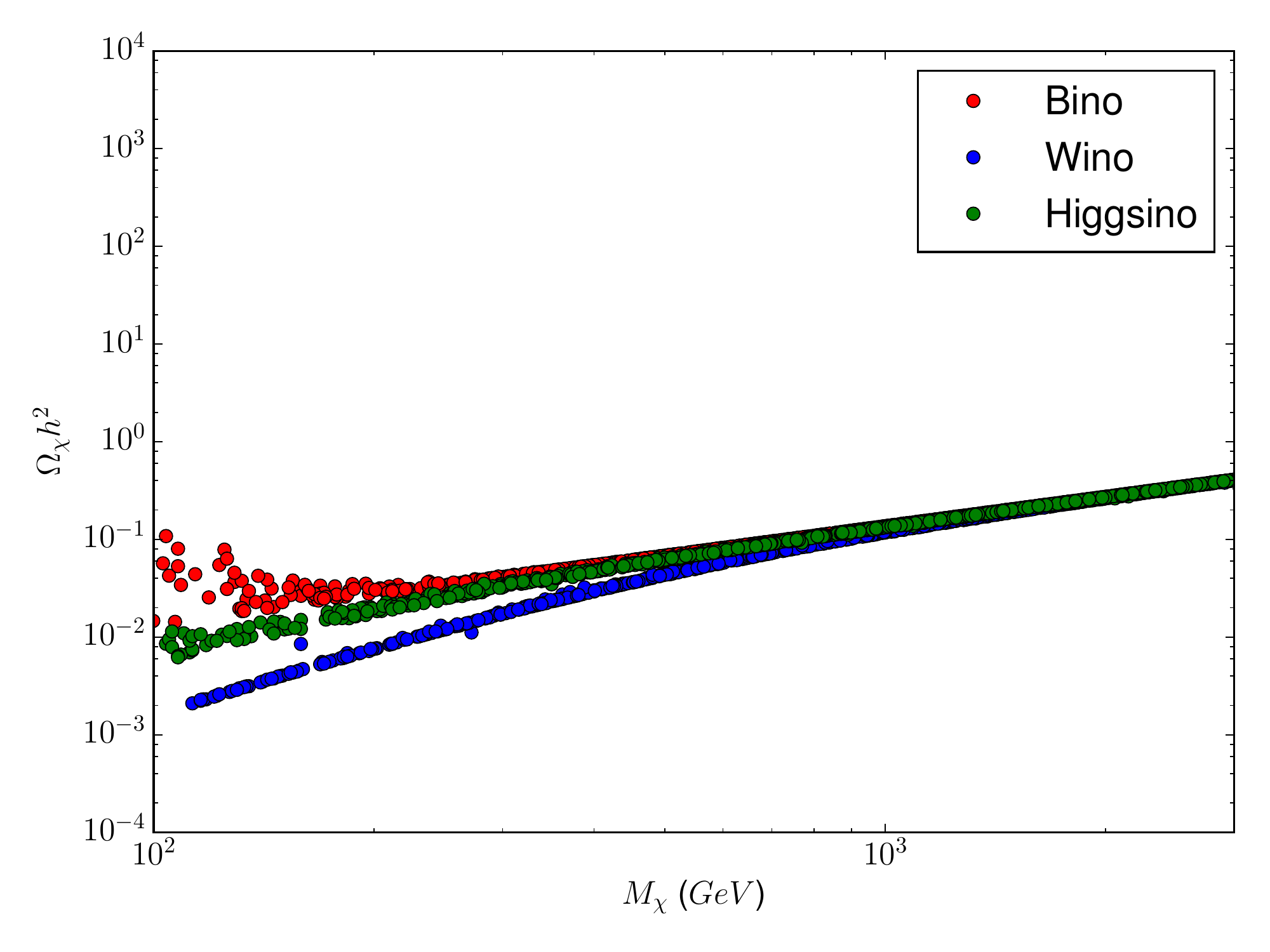} \\
\includegraphics[width=75mm]{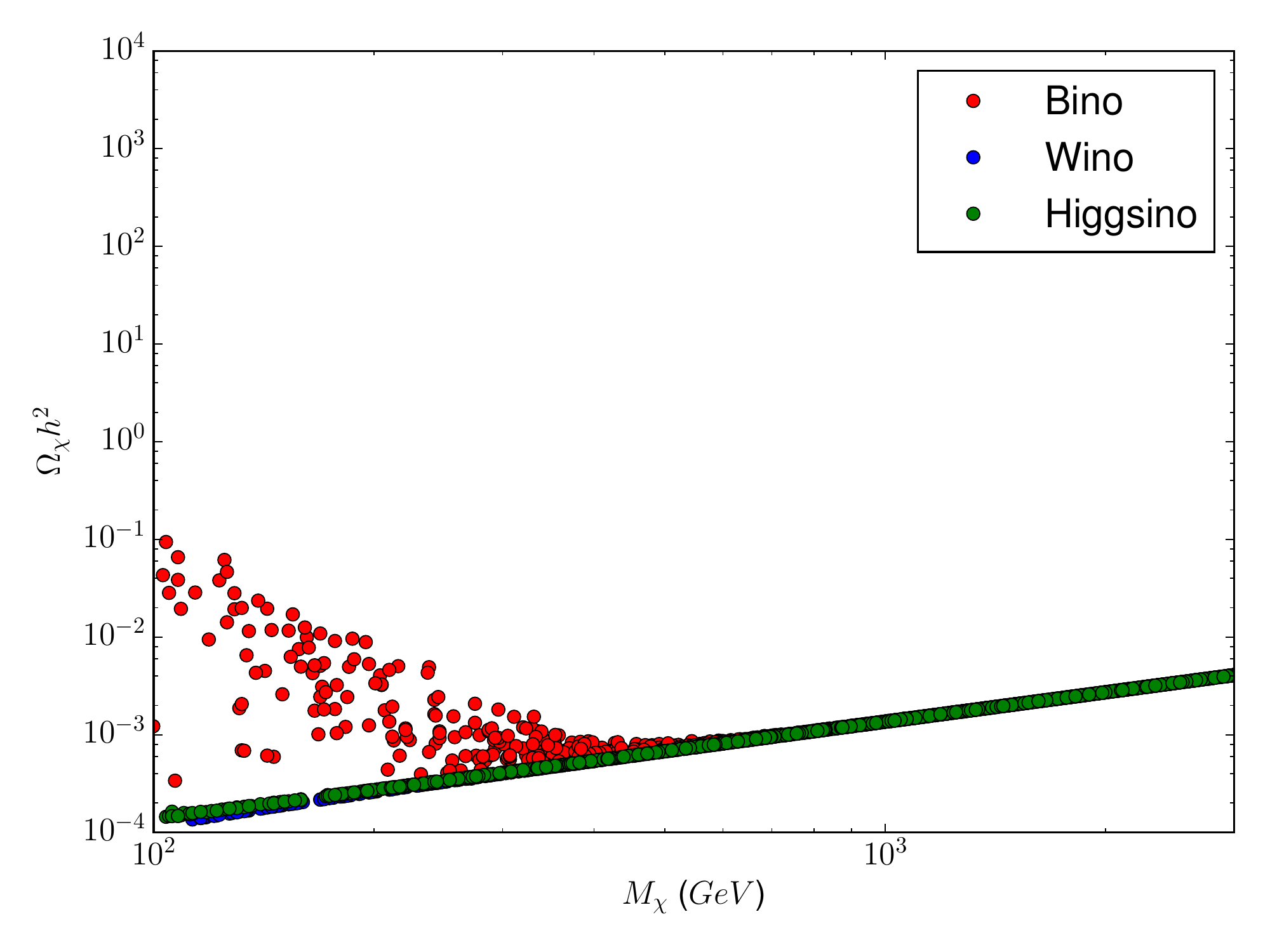}
\includegraphics[width=75mm]{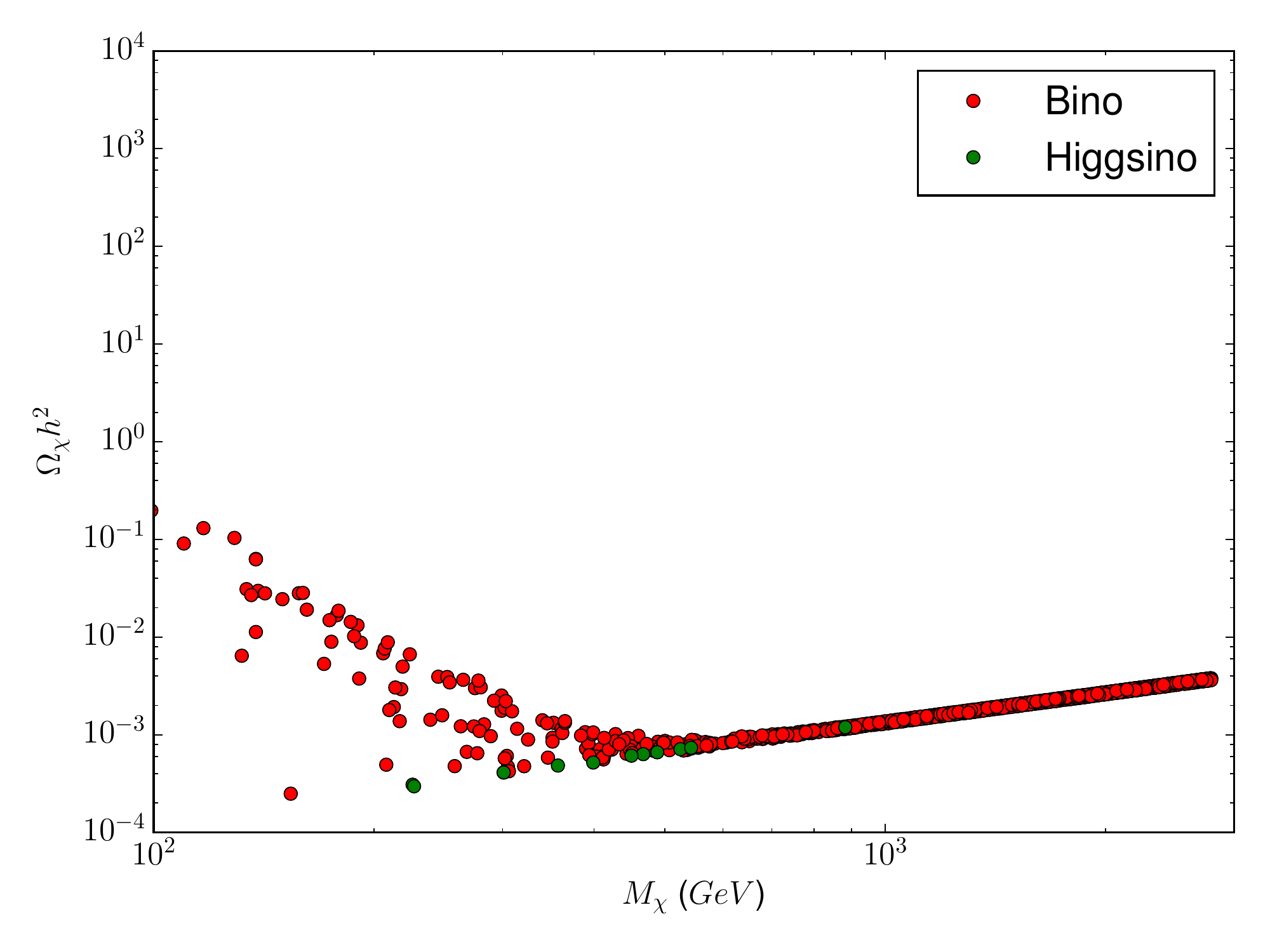}
\caption{The LSP relic density vs LSP mass for $M_\phi = 5 \cdot 10^6$
  GeV. The two upper frames are for the p10MSSM, with
  $B_\chi = 10^{-3}$ $(10^{-5})$ in the left (right) frame,
  respectively. The two lower frames are for $B_\chi = 10^{-7}$, in
  the p10MSSM (left) and cMSSM (right), respectively. The notation is
as in Fig.~\ref{5e5-pmssm}.
\label{5e6}}
\end{figure}

This is illustrated in Fig.~\ref{5e6}. The top--left frame is for
$B_{\chi} =10^{-3}$. We see that this relatively large branching ratio
leads to a sizable overdensity if the neutralino is bino--like. The
fact that most of the red points lie along a line implies that bino
annihilation is still insignificant in most cases. This implies that
the branching ratio (\ref{range_1}) can be interpreted as an {\em
  upper bound} for intermediate values of $M_\phi$, if the LSP is
bino--like.\footnote{For even larger $B_\chi$ bino annihilation might
  become significant; however, it cannot reduce the final relic
  density below that shown in the top--left frame of Fig.~\ref{5e6},
  which is already too large.} 

On the other hand, the annihilation cross sections for higgsino-- and
wino--like neutralinos with $M_\chi \lsim 1$ TeV are already so large
that eq.(\ref{decay1}) applies, where the final DM relic density is
independent of $B_\chi$ as long as $B_\chi > B_{\chi, \, {\rm
    crit}}$.
At the same time, the thermal production of higgsino-- and wino--like
LSPs still leads to a very small contribution to the final LSP relic
density. Note that for $M_\chi \sim 170$ GeV the green points in this
frame lie near the desired relic density. This allows to identify a
second region of parameter space that will give the correct DM
density, this time for higgsino--like LSPs:
\begin{equation} \label{range_2}
M_{\tilde H} \simeq 170 \ {\rm GeV} \cdot \left( \frac {M_\phi}
{5 \cdot 10^6 \ {\rm GeV}} \right)^{1/2}\ {\rm with} \ B_\chi \gsim 10^{-4} \cdot 
\left( \frac{5 \cdot 10^6 \ {\rm GeV}} {M_\phi} \right)\,.
\end{equation}
The bound on $B_\chi$ is a numerical approximation of the requirement
$B_\chi > B_{\chi, \, {\rm crit}}$, see eq.(\ref{B_crit}), for the range
of higgsino masses leading to approximately the correct DM relic density.
Of course, $M_\phi$ should not be so large that higgsinos thermalize at 
$T \simeq T_{\rm RH}$, i.e. the bound (\ref{mphi_max}) should also hold for
eq.(\ref{range_2}) to be applicable.

Comparing this figure with the left frame of
Fig.~\ref{thermal-relic-dm} we see that the gap between the blue and
green bands is even somewhat larger here than in standard
cosmology. As discussed earlier in this chapter, in both cases the
relic density scales like $1/\langle \sigma v \rangle$. However, in
standard cosmology it also scales linearly with $x_F = M_\chi / T_F$,
which grows logarithmically with increasing annihilation cross
section.  This factor, which does not exist in eq.(\ref{decay1}),
slightly reduces the difference of the predicted relic density of
wino-- and higgsino--like neutralinos in standard cosmology.

In the top--right frame we have reduced $B_\chi$ to $10^{-5}$. This
lies below $B_{\chi, \,{\rm crit}}$ for $M_\chi \gsim 1$ TeV in all
cases. In fact, eq.(\ref{range_1}) is still applicable in this region
of parameter space, but the required LSP mass near 1 TeV is already
outside the range we consider to be natural.

For slightly smaller $M_\chi$ the green and blue dots begin to diverge.
Here LSP annihilation after $\phi$ decay begins to be relevant, although
even at $M_\chi \simeq 100$ GeV the ratio between the green and blue points
is smaller than in the top--left frame, i.e. we have not fully reached the
regime described by eq.(\ref{decay1}). 

Moreover, bino--like LSPs, which have a much smaller annihilation
cross section, now begin to receive sizable contributions from the
thermal plasma. For this value of $M_\phi$ neutralinos with
$M_\chi \lsim 1$ TeV will always attain thermal equilibrium rather
early in the $\phi$ matter dominated epoch. For wino-- or
higgsino--like neutralinos this contribution to the final relic
density is still much smaller than the non--thermal contribution from
$\phi$ decays, even after late neutralino annihilation is
included. However, for bino--like neutralinos with very small
annihilation cross sections this thermal contribution can reproduce
the required relic density! This leads to another region of parameter space
with the required relic density, this time for bino--like states:
\begin{equation} \label{range_3}
M_{\tilde B} \sim 100 \ {\rm GeV} \cdot \left( \frac {M_\phi} 
{5 \cdot 10^6 \ {\rm GeV}} \right)^{3/2} \cdot \left( \frac
{10^{-13} \ {\rm GeV}^{-2} } { \langle \sigma v \rangle } \right)^{1/3}\,.
\end{equation}
This is a rough approximation. For example, we have ignored the factor
$x_{\rm FO}^4$ in eq.(\ref{mod_fo}); recall from the discussion of
that equation that $x_{\rm FO}$ has a stronger dependence on $M_\chi$
than the corresponding quantity in standard cosmology does. Moreover,
we have expressed the thermally averaged cross section in GeV units,
with $10^{-13}$ GeV$^{-2}$ being near the smallest cross section we
found for bino--like LSP. An increase of the annihilation cross
section can be compensated by a slight decrease of $M_{\tilde B}$, or
by an even smaller (relative) increase of
$M_\phi$. We should emphasize that eq.(\ref{range_3}) is applicable
only if $B_\chi$ is well below the value given in (\ref{range_1}). One
can also find combinations of parameters where
$\phi \rightarrow \tilde B$ decays and thermal $\tilde B$ production
both give comparable contributions to the final relic density; these
can be obtained quite easily from eqs.(\ref{range_1}) and
(\ref{range_3}).

In the bottom--left frame of fig.~\ref{5e6} we have reduced $B_\chi$
even further, to $10^{-7}$. This is well below
$B_{\chi, \, {\rm crit}}$ for all neutralinos, but the resulting
non--thermal contribution is still much larger than the thermal
contribution for higgsino-- or wino--like neutralinos. On the other
hand, the final relic density of bino--like neutralinos is now
dominated by the thermal contribution for $M_\chi \lsim 400$ GeV. This
contribution drops quite quickly with increasing $M_\chi$. This is
mostly from the explicit $M_\chi^{-3}$ factor in eq.(\ref{mod_fo});
the fact that the cross section also tends to increase with $M_\chi$
for bino--like LSP, as shown in Fig.~\ref{thermal-relic-dm}, also
contributes.  Finally, $x_{\rm FO}$ decreases logarithmically with
increasing $M_\chi$.  As expected, the red points near
$\Omega_\chi h^2 = 0.1$ that appeared in the top--right frame also
show up here, since for these points the non--thermal contribution was
negligible already in the former case, and is even smaller here.

The bottom--right frame of Fig.~\ref{5e6} is again for
$B_\chi = 10^{-7}$, but this time shows results for the cMSSM
scan. The red points here show a similar trend as in the bottom--left
frame, showing that the region approximated by eq.(\ref{range_3}) can
be accessed in the cMSSM as well.

\subsubsection{Heavy Moduli}

Finally, we consider a scenario with $M_\phi = 5 \cdot 10^7$ GeV, near
the upper end (\ref{mphi_max}) of the range where a $\phi$ dominated
epoch can modify the final LSP relic density. In fact, since now
$T_{\rm RH} = 25.49$ GeV, we expect that the neutralino relic density
should essentially coincide with that in standard cosmology if 
$M_\chi \lsim 300$ GeV.

\begin{figure} [h]
\includegraphics[width=75mm]{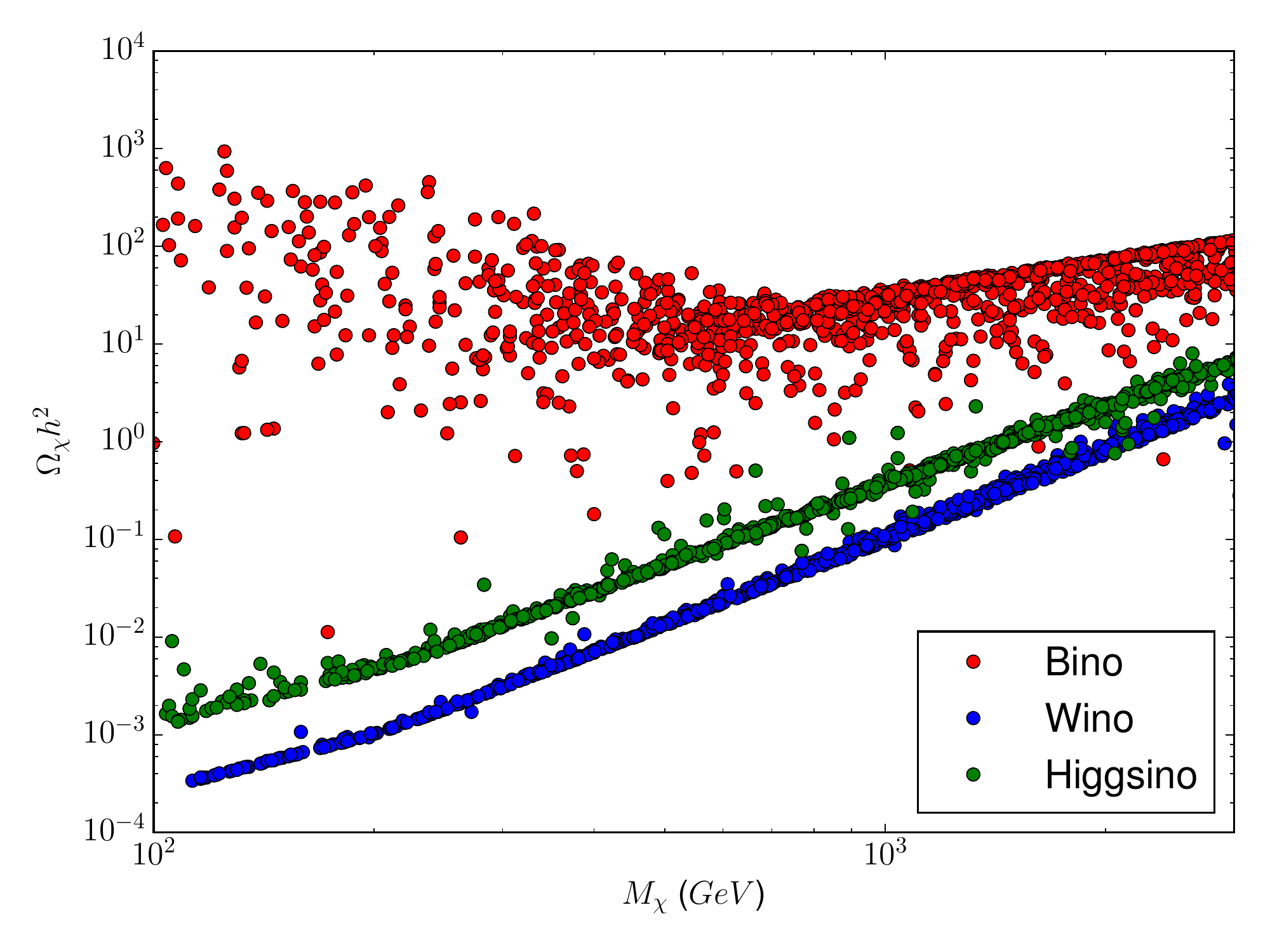}
\includegraphics[width=75mm]{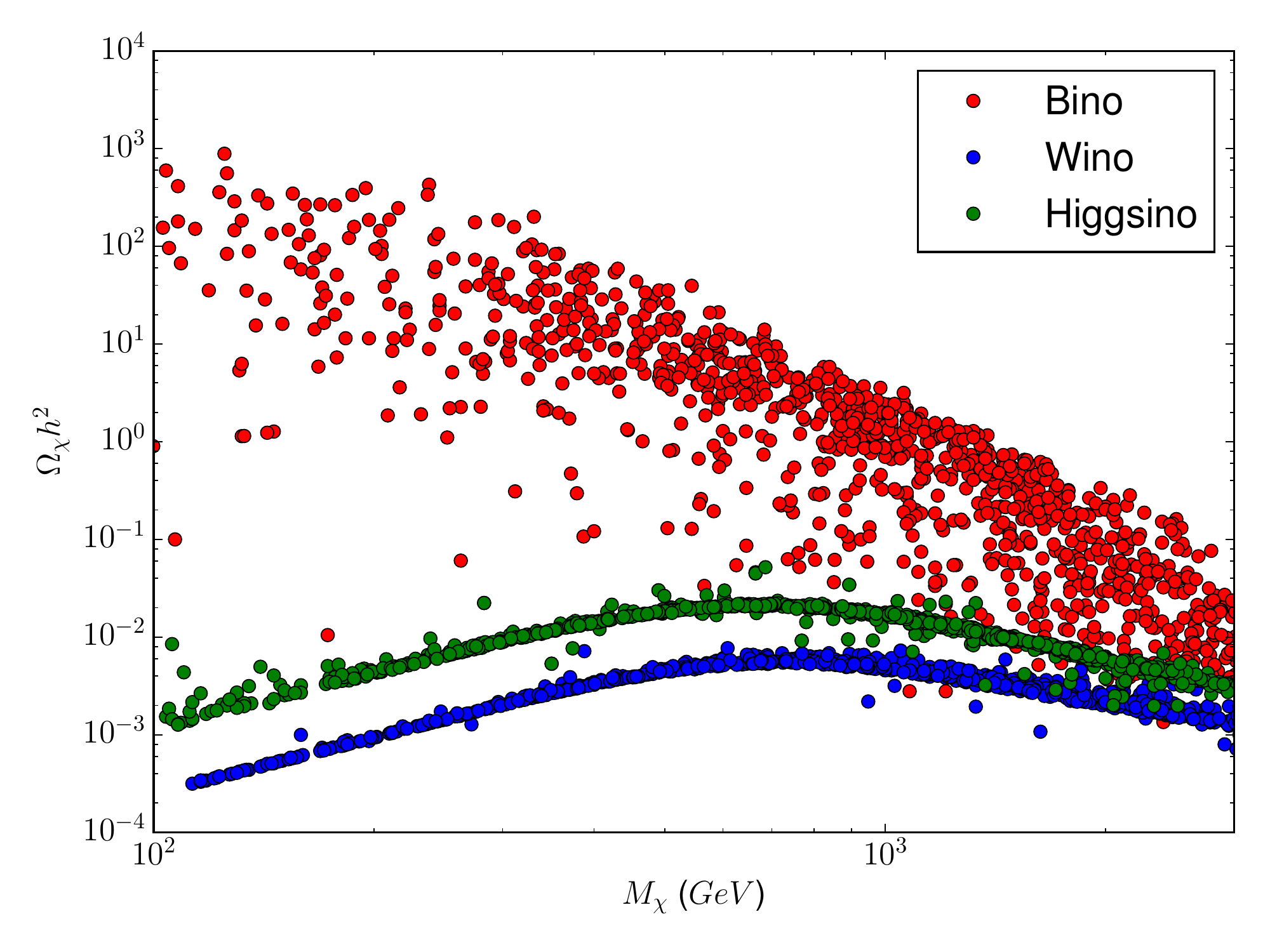}
\caption{The neutralino relic density vs neutralino mass for
$M_\phi = 5 \cdot 10^7$ GeV and $B_\chi = 10^{-3} \ (10^{-9})$ in the
left (right) frame. The notation is as in Fig.~\ref{5e5-pmssm}.
\label{5e7-pmssm}}
\end{figure}

This is borne out by Fig.~\ref{5e7-pmssm}. In the left frame we chose
$B_\chi = 10^{-3}$. This is well above $B_{\chi, \, {\rm crit}}$ for
both higgsino-- and wino--like neutralinos. Their relic density scales
like the inverse of the cross section, and hence essentially like
$M_\chi^2$. The slight change of slope of the lines of green and blue
points around $M_\chi \simeq 300$ GeV occurs because in standard
cosmology, which is applicable for $M_\chi$ below this point, the
explicit $x_F$ factor in the expression for the relic density slightly
softens the dependence on the cross section, and hence on $M_\chi$, as
remarked in the discussion of the top--left frame of
Fig.~\ref{5e6}. Note that thanks to this steepening of the slope,
higgsino--like neutralinos obtain the required relic density already
at $M_\chi \simeq 0.6$ TeV here, well below, and hence more natural
than, the value near $1$ TeV required in standard cosmology. In fact,
in this region of parameter space eq.(\ref{range_2}) is still
approximately applicable.

In contrast, nearly all red points indicating a bino--like neutralino
lie well above the required value, the exception being scenarios where
the effective annihilation cross section is enhanced by
co--annihilations or $s-$channel resonances. As expected, for
$M_\chi \lsim 300$ GeV the distribution of red points is very similar
to that in the left frame of Fig.~\ref{thermal-relic-dm}, which use
the same choices of MSSM parameters. The accumulation of red points
with $M_\chi \gsim 700$ GeV results from $\phi \rightarrow \tilde B$
decays, where late $\tilde B$ annihilation is still negligible, even
for these large densities, which are more than a factor of $100$ above
the desired value.

In the right frame of Fig.~\ref{5e7-pmssm} we therefore took a very
small branching ratio, $B_\chi = 10^{-9}$. The non--thermal
contribution to the neutralino relic density is then always
negligible. As expected, for $M_\chi \lsim 300$ GeV the green and blue
bands are the same as in the left frame. However, for $M_\chi \gsim 1$
TeV the relic density is essentially set by thermal freeze--out during
the $\phi$ matter dominated epoch.  Eq.(\ref{mod_fo}) shows that the
relic density there scales like
$1/(M_\chi^3 \langle \sigma v \rangle) \propto 1/M_\chi$, which
explains the fall--off with increasing $M_\chi$ at these large
neutralino masses. In between there is therefore a region where the
relic density depends only very weakly on the LSP mass, even though
LSP production is purely thermal here.  Unfortunately the value of the
relic density in this region of parameter space is about one order of
magnitude below the required value even for higgsino--like neutralino.

Most of the parameter points with bino--like LSP still predict a
significantly too large relic density. However, for $M_\chi \gsim 300$
GeV the upper envelope of the region populated by the red points
decreases quickly with increasing $M_\chi$; this is the same behavior
we saw in the two lower frames of Fig.~\ref{5e6}. As a result, for
$M_\chi \gsim 500$ GeV the density of red points with
$\Omega_\chi h^2 \simeq 0.12$ is significantly higher than in standard
cosmology, cf. Fig.~\ref{thermal-relic-dm}. The upper envelope even
approaches the desired value for $M_\chi \simeq 2.5$ TeV. However,
since these points violate our naturalness criterion, we do not
attempt to define another interesting region of parameter space in
which these points lie.

\section{Summary and Conclusions}
\label{sec:conclusion}

In this paper we investigated supersymmetric neutralino dark matter in
the framework of a non--standard cosmological scenario with an early
matter dominated epoch. Building on our earlier work
\cite{Drees:2017iod}, which improved the accuracy of the solution of
the relevant Boltzmann equations through a careful treatment of the
thermal medium, we looked for regions of parameter space where
relatively light neutralinos can form all of dark matter; our focus on
neutralinos with mass at or below 500 GeV is motivated by naturalness
arguments.

After setting up the basic framework, in Sec.~3 we reviewed neutralino
DM within standard cosmology. In agreement with many earlier studies,
we found that a bino--like neutralino typically has too high a relic
density, whereas higgsino-- or wino--like neutralinos can obtain the
desired relic density only for masses well above the ``natural''
range. Moreover, under the assumption that neutralinos form all of DM,
indirect searches exclude wino--like DM for masses below about $0.8$
TeV, which is already well above our naturalness cut--off. For
higgsino--like neutralinos the corresponding bound is around 400 GeV,
leaving some range of masses where higgsino DM could be (barely)
natural if it could get the required relic density. In contrast, if
the lightest neutralino is bino--like neither direct nor indirect DM
searches are very constraining \footnote{As pointed out in
 \cite{Erickcek:2015jza} and \cite{Erickcek:2015bda}, indirect bino signals can be 
boosted if the kinetic decoupling temperature $T_{\rm kd}$ is well above 
$T_{\rm RH}$. This is because the period of early matter domination will 
lead to an enhanced growth of structure at very small length scales \cite{Fan:2014zua}.
 These "microhalos" will be destroyed by free 
streaming of DM particles if $T_{\rm kd} \lsim T_{\rm RH}$. In
case of WIMPs, $T_{\rm kd} \gg T_{\rm RH} > 4$ MeV requires sfermion 
masses of tens of TeV, well above the range probed in our scan.}.

In Sec.~4 we therefore set out to find regions of parameter space
where a light bino, or a higgsino with mass near $400$ GeV, can obtain
the required relic density. The main new parameters, in addition to
those already present in standard cosmology, are the mass $M_\phi$ of
the heavy particle $\phi$ that accounts for the early
matter--dominated epoch, and its branching ratio $B_\chi$ into the DM
candidate. We found three distinct regions, described by
eqs.(\ref{range_1}), (\ref{range_2}) and (\ref{range_3}); this is the
main result of the present paper. In particular, for relatively 
small $M_\phi$, below $10^6$ GeV, neutralino annihilation is negligible,
and a simple relation for the required $B_\chi$ as a function of $M_\phi$
and $M_\chi$ results, see eq.(\ref{range_1}); this works for both bino--
and higgsino--like states. For higgsino--like neutralinos, annihilation
becomes important for $M_\phi > 10^6$ GeV, leading to a second range
where $\phi \rightarrow \tilde H$ decays followed by $\tilde H$ annihilation
produces the desired relic density; this is still a purely non--thermal
mechanism. Finally, for $M_\phi \gsim 5 \cdot 10^6$ GeV a third region
opens up, where bino--like neutralinos obtain the desired relic density
due to thermal freeze--out during the early matter dominated epoch,
see eq.(\ref{range_3}).

In this analysis we treated $M_\phi$ and $B_\chi$ as completely free
parameters, and assumed that the $\phi$ decay width scales like $M_\phi^3 /
M_{\rm Pl}^2$. In particular, in agreement with more general results found
in ref.\cite{Drees:2017iod}, we found that bino--like neutralinos can only have
the desired relic density even in this non--minimal cosmological scenario
if $B_\chi \lsim 10^{-4}$. It would be important to find concrete models
possessing a $\phi$ particle with the desired properties. In particular,
the upper bound on $B_\chi$ may not be easy to satisfy once higher--order
decays into three-- or even four--body final states have been included. We
leave this investigation to future work.

\acknowledgments We thank Alexander Pukhov for his help in modifying
{\tt{MicrOMEGAs}} code \cite{Belanger:2001fz}, and Giorgio Arcadi and
Hasan Serce for the clarification of their papers \cite{Arcadi:2011ev,
  Bae:2014rfa}. Finally, we thank Raghuveer Garani for useful
discussions. This research was supported by the Deutsche
Forschungsgemeinschaft via the TR33 ``The Dark Universe''. FH is
supported financially by the Deutsche Akademische Austauschdienst (DAAD). 
FH also thanks the hospitality and support of Goethe University of 
Frankfurt at the final stage of this work.

\vspace*{1cm}
{\bf Note Added:} While completing this work, ref.\cite{Arbey:2018uho}
appeared, which has some overlap with our work. Their results for a 
cosmology with early matter domination qualitatively agree with ours, 
but the main focus of their analysis is different.

\end{document}